\documentclass[10pt]{iopart}
\usepackage{fullpage}
\usepackage{amstext}
\usepackage{amssymb}
\usepackage{iopams}
\usepackage{graphicx}
\usepackage[dvips,colorlinks=true,urlcolor=purple,filecolor=purple,citecolor=purple,hyperfootnotes=true]{hyperref}
\usepackage[latin1]{inputenc}
\usepackage{xcolor}
\usepackage{ulem}

\newcommand{\grad}{\nabla}

\newcommand{\D}{{\mathcal D}} 
\newcommand{\HH}{{\mathcal H}} 
\newcommand{\bol}[1]{{\boldsymbol{#1}}} 
\newcommand{\smallmatrix}[1]{
  \text{\begin{footnotesize}
      $ \pmatrix{ %
        #1} $
    \end{footnotesize}}}

\newcommand{\ddp}[2]{\frac{\partial {#1}}{\partial {#2}}}

\renewcommand{\brace}[1]{{ \bol{#1}}}

\newcommand{\ket}[1]{|{#1}\rangle}

\newcommand{\eq}{{\textnormal{eq}}}   

\renewcommand{\em}{\it}
\begin{document}

\author{Julien Tailleur$^{1}$, Jorge Kurchan$^{2}$, Vivien Lecomte$^{3}$}

\address{\ $^1$ SUPA, School of Physics, University of Edinburgh,
  Mayfield Road, Edinburgh EH9 3JZ, Scotland}

\address{\ $^2$ Laboratoire PMMH (UMR 7636 CNRS, ESPCI, P6, P7),
         10 rue  Vauquelin, 75231 Paris cedex 05, France} 

\address{\ $^3$ DPMC, Universit\'e de Gen\`eve, 24, Quai Ernest Ansermet
  1211 Gen\`eve}

\title[Rare events in the hydrodynamic limit: mapping out of
  equilibrium back into equilibrium]{ Mapping out of equilibrium  into 
equilibrium in one-dimensional transport models. 
}

\date{\today}

\begin{abstract}
{Systems with conserved currents driven by reservoirs at the
 boundaries offer an opportunity for a general analytic study that is
 unparalleled in more general out of equilibrium systems. The
 evolution of coarse-grained variables is governed by stochastic {\em
   hydrodynamic} equations in the limit of small noise.}  As such it is
 amenable to a treatment formally equal to the semiclassical limit of
 quantum mechanics, which reduces the problem of finding the full
 distribution functions to the solution of a set of Hamiltonian
 equations. It is in general not possible to solve such equations
 explicitly, but for an interesting set of problems (driven Symmetric
 Exclusion Process and Kipnis-Marchioro-Presutti model) it can be
 done by a sequence of remarkable changes of variables. We show
 that at the bottom of this `miracle' is the surprising fact that
 these models can be taken through a non-local transformation into
 isolated systems satisfying detailed balance, with probability
 distribution given by the Gibbs-Boltzmann measure.  This procedure
 can in fact also be used to obtain an elegant solution of the much
 simpler problem of non-interacting particles diffusing in a
 one-dimensional potential, again using a transformation that maps the
 driven problem into an undriven one.

\end{abstract}


\maketitle

\section{Introduction}

Transport models are systems with conserved currents. In certain cases,
they are such that their evolution leads to equilibrium when they are
isolated or in contact with a thermal bath.  The probability
distribution is then of the Gibbs-Boltzmann form.  Coupling the
boundaries to several external sources may induce currents across the
bulk, driving the system out of equilibrium. In that case we do not
have any general explicit formula for the distribution of probability
of configurations, even in the stationary regime reached after long
times.

In order to make progress, one strategy has been to study systems
that, due to their specific symmetries, admit a complete
solution. Thus, in recent years a number of remarkable analytic
results have been found for simple transport models (see
\cite{Derrida2007} and references therein).  Two important examples are
the Simple Symmetric Exclusion Model (SSEP)--- a one-dimensional
system of particles, and the Kipnis-Marchioro-Presutti model (KMP)
\cite{KMP1982}, a model of energy transport.  For the former, Derrida,
Lebowitz and Speer~\cite{Derrida2001} (DLS) obtained an exact expression
for the large deviation function of density using a matrix method that
had been developed previously~\cite{Derrida1994}.

An alternative strategy is to restrict the calculation to the
probability distributions of coarse-grained variables.  If one
considers conserved quantities, then the macroscopic fluctuations obey
hydrodynamic equations with noise, the latter a manifestation of the
microscopic chaos or stochasticity.  Clearly, the more coarse-grained
the description, the lower the level of noise, because fluctuations
tend to average away.  One is thus lead, in the macroscopic limit, to
deterministic equations perturbed by stochastic terms whose variance
is of the order of the inverse of coarse-graining box size $N$.  As
usual, one can recast the problem in terms of the evolution in time of
the probability distribution. This is given by the Fokker-Planck
equation, which is closely analogous to a Schr\"odinger equation in
imaginary time, with the small parameter $N^{-1}$ playing the role of
$\hbar$.  The `semiclassical' treatment of these
equations~\cite{Freidlin1998} follows the same lines as the derivation
of classical from quantum mechanics, or geometric optics from wave
dynamics.  The logarithm of the transition probability obeys a
Hamilton-Jacobi equation whose characteristics are trajectories
satisfying Hamilton's equations.  Macroscopic Fluctuation theory,
developed by  Bertini, De Sole, Gabrielli, Jona-Lasinio
  and Landim (BDGJL)~\cite{Bertini2001,Bertini2005} is the
resulting classical Hamiltonian field theory describing coarse-grained
fluctuations.

Up to this point the formalism is completely general. However,
an analytic expression for the solutions of the classical
equations is not possible for every model, so that even in the coarse-grained
limit the problem has no closed solution.  Remarkably, for the
hydrodynamic limit of the driven SSEP~\cite{Bertini2001}, BDGJL
 were able to integrate explicitly the corresponding
Hamilton-Jacobi equations and recover the large-deviation function.
They thus followed a path that is in principle logically independent
of the one used to obtain the exact microscopic solution. Their
derivation amounts to rewriting the problem in a carefully chosen set
of variables.  An analogous strategy subsequently allowed Bertini,
Gabrielli and Lebowitz~\cite{Bertini2005} to do the same for the
hydrodynamic limit of KMP.

One is left wondering what is the underlying reason for the existence of
 changes of variable that allow to completely solve
 the Hamiltonian equations, and how general their applicability is.
  In this paper we show that {\em in the
 cases where this has been possible, there exists a non-local mapping
 taking the hydrodynamic equations of the model in contact with
 reservoirs into those of an isolated, equilibrium system.}  Large
 deviations and optimal trajectories are easily obtained in this
 representation using the detailed balance property, and can then be
 mapped back to the original setting in which detailed balance was
 broken by the boundary conditions. In the transformed, isolated model,
 spontaneous rare fluctuations are the time-reverse of relaxations to
 the average profile, but this symmetry is lost (as it should) in the
 mapping back to the original model. This accounts in this case for
 breaking of the Onsager-Machlup symmetry~\cite{Onsager1953} between
 birth and death of a fluctuation, which has received considerable
 interest~\cite{Luchinsky1997,Ciliberto2007} over the past few years.
A short account of this work has appeared elsewhere~\cite{Tailleur2007}.

The layout of this article is as follows. In section \ref{sec:PIROEP}
we review the expression of particle exclusion problem in terms of
spin operators.  We construct a (coherent-state) path integral for the
transition probability and derive from it the hydrodynamic limit. This route to the
 hydrodynamic limit is conceptually simple, and in addition
  has the advantage that the spin notation
makes  the symmetries and integrability properties of the
hydrodynamic limit explicit. Perhaps more surprisingly, the coherent-state
representation yields directly a set of (Doi-Peliti) variables in
terms of which the problem can be explicitly solved.

In section \ref{sec:FH} we take the more direct route starting
directly from Fluctuating Hydrodynamics, and deriving from it the
coarse-grained hydrodynamic equations. The reader may skip section
\ref{sec:PIROEP} and start here, the only loss is that the symmetries
of the hydrodynamic equations are not explicit and the boundary
conditions less straightforward.

In section \ref{sec:LDLNL} we review how the original stochastic
dynamics for a single density field $\rho(x)$ leads, in the low-noise
limit, to a classical Hamiltonian field theory in a phase-space with {\em two}
fields $\rho(x)$ and $\hat \rho(x)$. The large deviation function --
the logarithm of the probability of a configuration in the stationary
regime -- is given by the action of a `classical' trajectory
$[\rho(x,t), \hat \rho(x,t)]$ reaching the configuration at very long
times. The formalism is a variant of the low-noise
Freidlin-Wentzell formalism, itself an implementation of the usual WKB
semiclassical theory.  

For an undriven system with Detailed Balance
(section \ref{sec:DBR}) the dynamics has a symmetry between `downhill'
relaxations and `uphill' excursions.  Using this symmetry one obtains the uphill
trajectories as the time-reversed of the downhill ones, and this allows 
to compute 
explicitly the large-deviation function.  In
the case of driven transport models, the boundary terms violate
detailed balance, and there is no obvious symmetry playing the role 
of time-reversal,  no general  method to find the uphill trajectories, and hence
no explicit solution for the large-deviation function.

In section \ref{sec:FetHatF}, a paraphrase of the solution of 
BDGJL,  we show that in the particular case of
the SSEP, there is a very special set of variables that allows to
solve completely the driven problem.
The main result of this paper, in section
\ref{sec:BackToEq}, is to show  that at the bottom of this possibility is the fact
that there is a non-local mapping converting the driven chain
into an isolated, undriven one.  In section \ref{sec:KMP} we briefly
show that one can apply the same arguments to solve the KMP model.

To conclude, in section \ref{sec:duality_FP} we discuss how the method can be
  applied to the much simpler case of non-interacting
particles diffusing in a generic  potential,
  driven out-of-equilibrium by boundary terms.

\subsection{Mapping a driven into an undriven problem}

Let us consider here the simplest example of a mapping from out of
equilibrium to equilibrium. This exercise will help to fix ideas, and
to see how much one gets out of such an approach.

Consider a diffusion process given by a Fokker-Planck equation driven
by boundary conditions $P(0)=p_o$ and $P(L)=p_L$
\begin{equation}
  \label{eq:FPcontinuum}
  \dot P= \frac{d}{dx} \left(T\frac{d}{dx}+\frac{dV}{dx}\right)P
\end{equation}
Because of the boundary conditions, the current
$J=\left(-T\frac{d}{dx}-\frac{dV}{dx}\right)P$ is in general non-zero.

Now let us {introduce} $e^{\beta V}P=P_1$. {Equation
  \eref{eq:FPcontinuum} maps to the backward Fokker-Planck equation}:
\begin{equation}
  \dot P_1=  \left(T \frac{d}{dx}-\frac{dV}{dx}\right) \frac{d}{dx} P_1
\end{equation}
 Defining  $P'=\frac{d}{dx} P_1$, we get:
\begin{equation}
  \dot P'= \frac{d}{dx} \left(T\frac{d}{dx}-\frac{dV}{dx}\right)P'
  \label{eq:evolPprimed_FP}
\end{equation}
This takes the form of a probability $P'$ describing an evolution in a
potential $-V$. {The remarkable fact is that the original boundary
  condition on $P$ expresses in the new variables just a normalization
  for $P'$:}
\begin{equation}
  \int_0^L P' dx =   {e^{\beta V(L)}p_L-  e^{\beta V(0)}p_o}
\end{equation}
Furthermore, it also implies that the current $J'$
associated with $P'$
vanishes at the ends $\Lambda=(0,L)$, since:
\begin{equation}
  \dot P(\Lambda)=0 \rightarrow \frac{d}{dx}
  \left(T\frac{d}{dx}+\frac{dV}{dx}\right)P(\Lambda)=0 \rightarrow
  \left(T\frac{d}{dx}-\frac{dV}{dx}\right)P'(\Lambda)=-J'(\Lambda)=0
  \label{jj}
\end{equation}
{\em $P'$ thus evolves with a Fokker-Planck equation with potential
  $-V$ and no current at the boundaries; the integral of $P'$ is 
  conserved and the process satisfies detailed balance.}  The stationary
measure is thus $ P_{stat}'(x) \propto e^{\beta V(x)}$
\begin{equation}
  { P_{stat}' \propto e^{\beta V} \rightarrow P_{stat}(x) = p_0 e^{-\beta
      [V(x)-V(0)]} +c \int_0^x e^{-\beta[V(x)-V(x')]}dx'}
\end{equation}
where the constant $c$ is fixed by the right boundary condition
$P(L)=p_L$. The overall distribution then reads
\begin{equation}
  { P_{stat}(x)=\frac{p_0 e^{-\beta[V(x)-V(0)]}\int_x^L e^{\beta
        V(x')}dx'+p_L e^{-\beta[V(x)-V(L)]}\int_0^x e^{\beta
        V(x')}dx'}{\int_0^Le^{\beta V(x')}dx'}}
  \label{eq:Pstat_FP}
\end{equation}
This result could have been obtained from the beginning by
quadratures~\cite{Ledoussal1995}. Actually, we have obtained much
more, {since we have mapped the evolution operator into the one of
  an equilibrium problem} and we now understand the time-evolution of
the primed variable as the relaxation of an isolated system: even
without calculating anything, we have an intuition of the qualitative
behavior.

To see that the transformation is non-local, we may consider the
expectation value of a function $O(x)$: {
\begin{eqnarray}
  \langle O\rangle(t)&=& \int_0^L dx \; O(x) P(x) = \int_0^L dx \;
  O(x) e^{-\beta V(x)}P_1(x)\nonumber \\
&=& \int_0^L dx \; O'(x) P'(x) \; - \;\;
  O'(L) p_L e^{\beta V(L)}
\end{eqnarray}}
All the time dependence is given by the expectation value
of the new operator:
\begin{equation}
O'(x) \equiv - \int_0^x dy \; O(y) e^{-\beta V(y)}
\end{equation}
which is a non-local function of the original one.

In section \ref{sec:duality_FP} we shall show that actually one can
use the same non-local transformation to map {\em any} single-particle
diffusion model in a one-dimensional potential with sources at the
ends into the same {equilibrium} diffusion problem with no sources.
 
In the rest of the paper, we shall meet an analogous situation, for
models with many interacting particles. The main difference is that we
shall be making {\em a transformation at the level of fields, not of
  their probability} ($P$ here is a probability of the $0$-dimensional
`field' $x$).



\pagebreak

\section{Path-integral representations of exclusion processes}
\label{sec:PIROEP}
In this section we introduce the exclusion processes. We write their
evolution matrix in terms of (quantum) spin operators. This makes
explicit the fact that the problem has more symmetries than mere
particle conservation. Using {standard} spin coherent state
techniques we present {a novel derivation} of the path integral
and then take the hydrodynamic limit. We also obtain a natural set of
variables $F,\hat F$ (related to the stereographic representation of
the spins) which are often implicitly used in the
literature. Finally, we write the hydrodynamic equations in terms of
the average particle density $\rho$, and its conjugate variable $\hat
\rho$.

{The main result of this section is the construction of an
  hydrodynamic action which gives the logarithm of the transition
  probability between two smooth density profiles, together with the
  associated spatio-temporal boundary conditions. It is given by
  equations (\ref{eqn:actiondyn}-\ref{eqn:BC2}).} The reader familiar with
these representations of exclusion processes may skip this section,
and find the more standard approach in the next one.

\subsection{Partial Exclusion Process}

Let us consider the symmetric partial exclusion process, a
generalization of simple symmetric exclusion processes introduced in
\cite{Schutz1994}. It consists of a one dimensional lattice gas, for
which all sites can be occupied by at most $2j$ particles. The
probability of a jump between a site and its neighbor is proportional
to both the occupation number of the starting site and the
{proportion} of vacancies of the target one\footnote{{The prefactor
    $1/(2j)$ ensures that the rate at which a fixed number of
    particles jump to an empty site does not diverge with $j$.}}:
\begin{eqnarray}
  \label{ratedefinition}
  W(\dots,n_k,n_{k+1},\dots \to\dots, n_k+1,n_{k+1}-1,\dots)= \frac
  p{2j}\, (2j-n_k)\, n_{k+1}\\ W(\dots,n_k,n_{k+1},\dots \to
  \dots,n_k-1,n_{k+1}+1,\dots)= \frac p{2j}\, n_k\,
  (2j-n_{k+1})\nonumber
\end{eqnarray}
We fix a time-scale by choosing $p=1/2$. The system can be put in
contact at sites $1$ and $L$ with reservoirs of densities $\rho_0$ and
$\rho_1$. This is usually done by introducing four rates
$\alpha,\delta$ and $\gamma,\beta$ which correspond to deposition and
evaporation of particles at site $1$ and $L$, respectively. We thus
have the added rates of interchange with the reservoirs
\begin{eqnarray}
  \label{ratedefinitionBC}
  W(n_1,\dots \to n_1+1,\dots)&=  {\alpha}\, (2j-n_1)\qquad
  &W(\dots,n_L \to \dots,n_L+1)= { \delta}\,
  (2j-n_L)\nonumber\\ W(n_1,\dots \to n_1-1,\dots)&= {
  \gamma}\, n_1\qquad &W(\dots,n_L \to\dots,n_L-1)= 
  {\beta}\, n_L
\end{eqnarray}

\begin{figure}
  \begin{center}
    \includegraphics[clip]{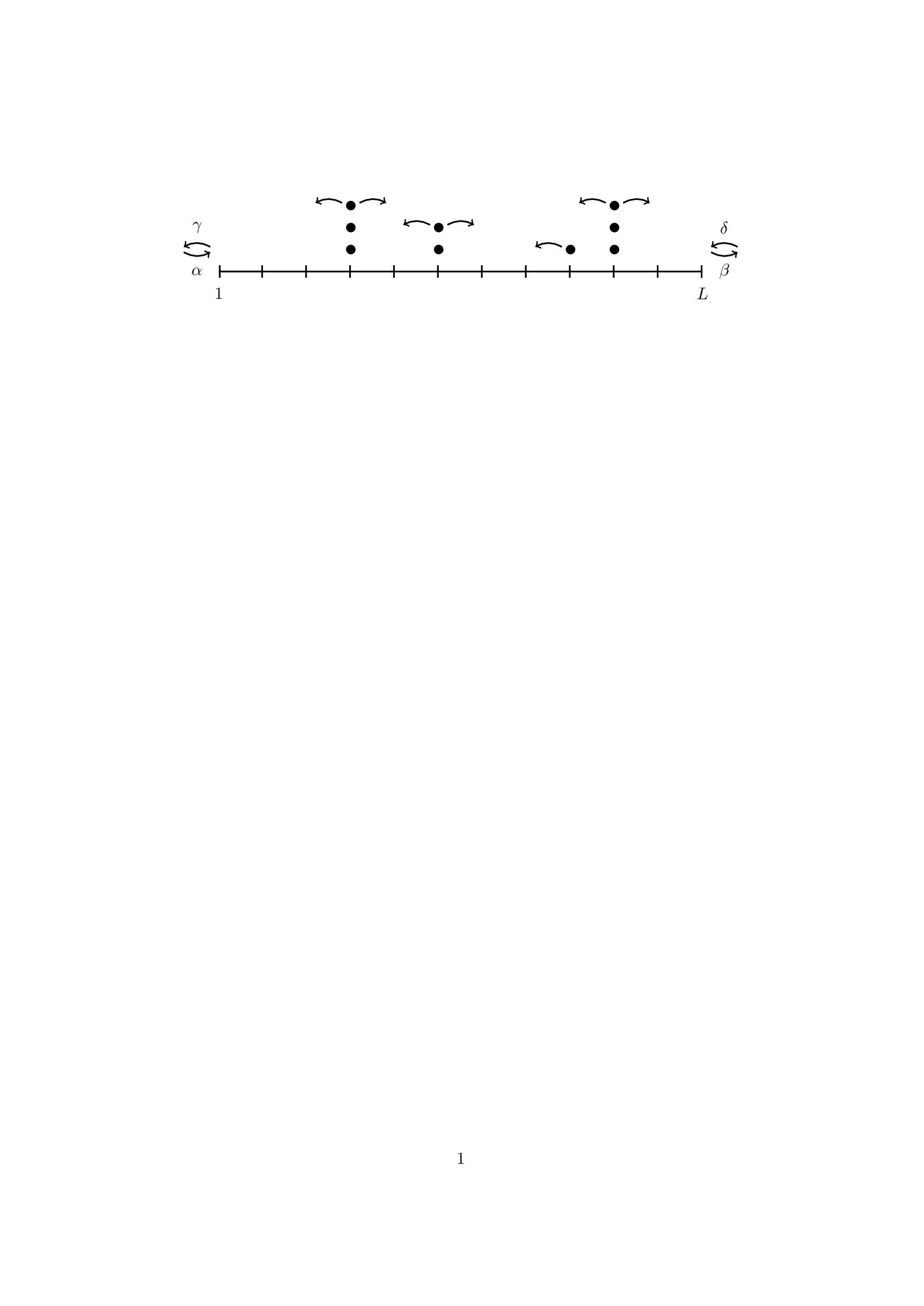}
    \caption{Schematic representation of a partial exclusion process
      for j=3/2. There can be at most 3 particles per site. Particles
      are injected at site $1$ and $L$ with rate $\alpha$ and $\delta$
      and can jump to the corresponding reservoirs with rate $\delta$
      and $\beta$.}
    \label{fig:PEP}
  \end{center}
\end{figure}

Though the bulk-diffusion is symmetric, and the system thus satisfies
a local detail balance relation, it can be driven out of equilibrium
by the boundaries, if the densities imposed by the reservoirs are different:
\begin{equation}
  \rho_0=\frac{\alpha}{\alpha+\gamma}\neq \rho_1=\frac{\delta}{\delta+\beta}
\end{equation}
These models are amongst the simplest interacting many particle
systems driven out of equilibrium by the boundary sources. A schematic
representation of the partial exclusion process is shown on figure
\ref{fig:PEP}.
For $j=1/2$, we recover the usual SSEP, which is known to be related
to the $1/2$ representation of the $SU(2)$ group, whereas the partial
exclusion processes correspond to the spin $j$
representation~\cite{Schutz1994}. For sake of completeness, we give
the details of the relations with the spin operators in the next
subsections and use the $SU(2)$ coherent states to construct a
path-integral representation afterwards.
\subsection{Master equation and spin representation}
\label{sec:mastereqandspin}
The evolution of the probability $P({\boldsymbol n})$ of observing a
configuration defined by the vector of occupation number{s}
${\boldsymbol n} = (n_1, \dots, n_L)$ is given by the master equation
\begin{equation}
  \ddp{P(\brace{n})}{t}=\sum_{\brace{n'}\neq \brace{n}} W(\brace{n'}
  \to \brace{{n}}) P({\boldsymbol n'}) -
  W(\brace{n}\to\brace{n'})P(\brace{n})
\end{equation}
where $W(\brace{n'} \to \brace{{n}})$ is the transition rate from
configuration $\brace{n'}$ to configuration $\brace{n}$. To keep the
notation as compact as possible, we introduce $n_k^+=n_k+1$ and
$n_k^-=n_k-1$. From \eref{ratedefinition} and \eref{ratedefinitionBC},
the master equation reads
\begin{eqnarray}
  \frac{\partial P({\boldsymbol n})}{\partial t} &=
\frac 1 {4j}\sum_{k=1}^{L-1}
  \Big[(2j-n_k^-) n_{k+1}^+ P(\dots,n_k^-,n_{k+1}^+,\dots)+
   n_k^+ (2j-n_{k+1}^-) P(\dots,{n_k^+,n_{k+1}^-},\dots)\nonumber\\
& - [  (2j-n_k)n_{k+1}
  +  n_k (2j-n_{k+1})]P(\dots,n_k,n_{k+1},\dots) \Big]\label{eq:MeqPEP}\\
&+\alpha(2j-n_1^-)P({n_1^-},\dots)
 +\gamma n_1^+ P({n_1^+},\dots)
 +\delta (2j-n_L^-) P(\dots,{n_L^-})
+\beta n_L^+ P(\dots,{n_L^+})\nonumber\\
 &-[\gamma n_1 +\alpha (2j-n_1) + \delta (2j-n_L) +\beta n_L] P(n_1,\dots,n_L)\nonumber
\end{eqnarray}
The two first lines correspond to the dynamics in the bulk whereas the
last ones stand for the interaction with the reservoirs.

Let us now introduce a non-Hermitian representation of the SU(2)
{group} to write the master equation in an operatorial form. A
configuration of the system can be written as the tensor product of
 states of each sites $i$: $\ket{\psi}=\otimes_i \ket{\psi_i}$,
where each state $\ket{\psi_i}$ is given by a $2j+1$ components
vector, such that an occupation number equal to $n$ is represented by
\begin{equation}
  \ket{n}=\smallmatrix{0\cr\vdots\cr 0\cr 1\cr 0\cr \vdots \cr 0}
\end{equation}
where the $1$ is on the $(n+1)$th line, starting from the top. One 
then defines the $(2j+1)\times(2j+1)$ matrices:
\begin{equation} \label{eq:spin_matrx}
  S^+=  \smallmatrix{
    0&\dots&\dots&\dots&0             \cr
    2j&\ddots&&&\vdots\cr
    0&\ddots&\ddots&&\vdots \cr
    \vdots&\ddots&\ddots&\ddots&\vdots\cr
    0&\dots&0&1&0
  }
  \quad
  S^-=\smallmatrix{
    0&1&0&\dots         &0   \cr
    \vdots&\ddots& \ddots &\ddots&\vdots     \cr
    \vdots&&\ddots&\ddots&0 \cr
    \vdots&&&\ddots&2j\cr 
    0&\dots&\dots& \dots&0    \cr
  }
  \quad
  S^z=
  \smallmatrix{
    -j&0&\dots&\dots&0           \cr
    0&\ddots&\ddots&&\vdots    \cr 
    \vdots    &\ddots&\ddots&\ddots&\vdots \cr
    \vdots&&\ddots& \ddots&0    \cr
    0    &\dots&\dots&0&j
  }
\end{equation}
whose action on the state $|n_i\rangle$ are given by
\begin{eqnarray}
  \label{eqn:sigma}
  S_i^+ \ket{n_i}=(2j-n_i) \ket{n_i+1}\ \qquad
  S_i^- \ket{n_i}=n_i \ket{n_i-1} \qquad
  S_i^z \ket{n_i}=(n_i-j) \ket{n_i}\\
  S_i^x = \frac 1 2 [ S_i^++S_i^-]\qquad S_i^y=\frac 1 {2\rmi} [S_i^+ - S_i^-]\nonumber
\end{eqnarray}
Direct computations show that they satisfy the commutation relations
\begin{equation}
  \label{eq:CommutQuantumSpins}
        [S^z,S^\pm]=\pm S^\pm  \qquad
  [S^+,S^-]=2S^z\qquad [S^i,S^j]=\rmi\,\epsilon^{ijk}\, S^k\quad \forall\, i,j,k\,\in \{x,y,z\}
\end{equation}
They thus form a $2j+1$ dimensional representation of the SU(2)
{group}, that is a representation of spin $j$, the usual magnetic
number $m_i$ being related to the occupation number $n_i$ through
$m_i=n_i-j$.  This is a non-unitary representation as $S^+$ is not the
adjoint of $S^-$~\footnote{Note that a similar path could be followed
  with the usual unitary representation of SU(2), see
  \ref{app:UnitorNonUnit}. It would however lead to additional
  temporal boundary terms in the path integral, which makes the
  analysis of the action less straightforward.}.

The evolution operator of the partial exclusion process can now be
written using these spin {operators}. To make this explicit, let us
introduce the vector $\ket{\psi}=\sum_{\brace{n}} P(\brace{n})
\ket{\boldsymbol n}$ where the sum runs over all the possible
configurations ${\boldsymbol n}=(n_1,\dots,n_L)$. Using the master
equation, the time derivative of $|\psi\rangle$ is given in terms of
the spin operators by
\begin{eqnarray}
  &\ddp{\ket{\psi}}{t}=-\hat H  \ket{\psi};\qquad \hat H = \hat H_1+\hat H_B+\hat H_L; \nonumber\\
  \hat H_B &= \frac 1 {4j}\sum_{k=1}^{L-1} 
  \left[- S_{k+1}^- S_k^+ + (j+S_{k+1}^z)(j-S_{k}^z)\right]
  +\left[-S_{k+1}^+ S_k^- + (j-S_{k+1}^z)(j+S_{k}^z)\right];\nonumber\\
  \hat H_1&=-\alpha\left[S_1^+-(j-S_1^z)\right]-\gamma\left[S_1^--(j+S_1^z)\right];\nonumber\\
  \hat H_L&=-\delta\left[S_L^+-(j-S_L^z)\right] -\beta\left[S_L^--(j+S_L^z)\right].\nonumber\\
\end{eqnarray}
$\hat H_B$ corresponds to the dynamics in the bulk, whereas $\hat
H_1$ and $\hat H_L$ result from the coupling with the reservoirs on
the sites $1$ and $L$. Introducing spin vectors $\vec S_k
=(S_k^x,S_k^y,S_k^z)$, $\hat H_B$ can be written in a more compact way as
\begin{equation}
  \label{eqn:hamspindiscret}
  \hat H_B=-\frac{1}{2j}\sum_{i=1}^{L-1}  \left(\vec S_k \cdot \vec S_{k+1}-j^2\right),
\end{equation}
 This is the usual connexion between exclusion processes and spin
chains~\cite{Dhar1987,Gwa1992,Schutz1994,Fogedby1995}.

\subsection{Path integral representation in the hydrodynamic limit}
\label{sec:spin_coh_st}
The hydrodynamic limit of lattice-gas models is usually achieved
through the definition of a coarse-grained density field $\rho$,
averaged over macroscopic boxes, the  size of which is then sent to
infinity~\cite{Spohn1983}. Here, each site of the lattice already
contains up to $2j$ particles.
 We can thus obtain a hydrodynamic limit by taking the limit of
$j,L$ going to infinity.
In the macroscopic limit, all the spin
representations are equivalent, $j$ and $L$ enter through the
combination $jL$.  This is just the manifestation of the fact
\cite{Fogedby1995} that the hydrodynamic variables represent the total
spin in a box, whatever the spin of the elementary sites.

\subsubsection{Coherent states.}
To construct a path integral representation of $\hat H$, we shall use,
for each site $k$ of the lattice, the following right and left spin
coherent states \cite{Perelomov1986}
\begin{eqnarray}
  \label{eq:def_spin_coh_st}
  |z_k\rangle &= \frac{1}{(1+\bar z_k z_k)^{j}}\,\rme^{z_k S_k^+}
  |0_k\rangle= \frac{1}{(1+\bar z_k z_k)^{j}} \sum_{0\leq n_k\leq 2j}
  \smallmatrix{ 2j\cr n_k} z_k^{n_k} |n_k\rangle,\nonumber\\
 \langle z_k| &= \frac{1}{(1+\bar z_k z_k)^{j}}\,\langle 0_k|\,
 \rme^{\bar z_k S_k^-}= \frac{1}{(1+\bar z_k z_k)^{j}} \sum_{0\leq
 n_k\leq 2j} \langle n_k| \,\bar z_k^{n_k}.
\end{eqnarray}
For an extended system of $k$ sites, one  introduces the tensor
product
\begin{equation}
  |{\bol z} \rangle = \bigotimes_k |z_k \rangle.
\end{equation}

Due to the non-Hermiticity of the representation \eref{eq:spin_matrx},
$|z_k\rangle$ is not the adjoint of $\langle z_k|$. The construction
of the path integral relies on the following representation of the
identity
\begin{equation}
  \label{eq:represid}
  \int \rmd\mu(z_k) \, |z_k\rangle\langle z_k| = \hat 1
  \quad\textnormal{with}\quad
  d\mu(z_k)=\frac{2j+1}{\pi} \frac{d^2z_k}{(1+z_k\bar z_k)^2}.
\end{equation}

\subsubsection{Action functional in the large spin limit.}
\label{sec:action_func}

The functional approach for spin operators has some subtleties~\cite{Garg2000}
(see Solari~\cite{Solari1987}, Kochetov~\cite{Kochetov1995}, Vieira and
Sacramento~\cite{Vieira1995} for a derivation of the
action and of the associated time boundary conditions). For sake of
clarity, most of the technical details are presented in
\ref{app:action_func} and we simply outline below the main steps in
the path-integral construction.

{Keeping in mind that we ultimately want to describe a continuum
theory, we replace number occupations by `discrete' densities:
\begin{equation}
  \rho_k=\frac{n_k}{2j}
\end{equation}}
We would like to compute $P(\bol{ \rho^f},T;\bol{ \rho^i},0)$, the
probability of observing the system in state $(\rho_1^f,\dots,\rho_L^f)$ at
time $T$, starting from $(\rho_1^i,\dots,\rho_L^i)$ at time $0$, that is the
propagator between two states $\langle \bol{ \rho^f}|$ and $|\bol{ \rho^i}
\rangle$ with fixed initial and final number of particles in each
site. Using $2L$ representations of the identity~\eref{eq:represid}, we
write
\begin{eqnarray}
  \label{eqn:physicalpropagator}
  P(\bol{ \rho^f},T;\bol{ \rho^i},0)
  &=\langle \bol{ \rho^f}| \rme^{-T \hat H}| \bol{ \rho^i}\rangle 
  =\int\prod_k \rmd\mu(z_k^f)\rmd\mu(z_k^i)\:
  \langle \bol{ \rho^f}|\bol{ z^f}\rangle 
  \langle \bol{ z^f}| \rme^{-T \hat H}| \bol{ z^i}\rangle
  \langle \bol{ z^i}|\bol{ \rho^i}\rangle 
\end{eqnarray}
 Because we are  interested in the
hydrodynamic limit (large $jL$), we may first take a  large $j$ limit
and then send $L$ to $\infty$. In this `large spin' limit,
\eref{eqn:physicalpropagator} reads (see details in
\ref{app:action_func})
\begin{eqnarray}
  \label{eq:propninf}
  P(\bol{ \rho^f},T;\bol{ \rho^i},0)= \int\prod_k
  \rmd\mu(z_k^f)\rmd\mu(z_k^i)\int\D\bol{ \bar z}\D \bol{ z}
  \prod_{e=i,f}\delta\left( \frac{\bar z_k^ez_k^e}{1+\bar
  z_k^ez_k^e}-\rho_k^e\right)\, \exp[-S],\\
  S= 2j \sum_k \left[\frac{z_k \bar
  z_k}{1+z_k \bar z_k}\log\bar z_k - \log(1+z_k \bar z_k)\right]_i^f
  +2 j \int \rmd t \left[\sum_k \frac{\bar z_k \dot z_k}{1+\bar z_k
  z_k} -\HH(\bar z,z)\right]
\end{eqnarray}
The role of the Hamiltonian $\HH(\bol{ \bar z},\bol{ z})$ is played by
the quantity $ -\frac 1{2j} \langle \bol{ z}|\hat H | \bol{ z}\rangle$
which is computed using~\cite{Garg2000}:
\begin{equation} 
  \label{eq:classicS} 
\langle z_k | S_k^+ | z_k\rangle
= 2j \frac{\bar z_k}{1+z_k\bar z_k}; \qquad 
\langle z_k | S_k^- |
  z_k\rangle = 2j \frac{ z_k}{1+z_k\bar z_k}; \qquad 
\langle z_k | S_k^z
  | z_k\rangle =j \frac{z_k\bar z_k -1 }{z_k\bar z_k+1} \nonumber
\end{equation}
Explicitly, this yields
\begin{eqnarray}
  \label{eqn:hamiltonianz}
  \HH(\bar z,z)= \HH_B(\bar z,z)+\HH_0(\bar z,z)+\HH_L(\bar
  z,z)\\
  \HH_B(\bar z,z)=- \sum_{k=1}^{L-1} \frac{ z_k}{1+\bar z_k z_k}
  \frac{ z_{k+1}}{1+\bar z_{k+1} z_{k+1}} \frac{(\bar z_{k+1}-\bar
    z_k)^2}2 + \left(\frac { z_{k+1}} {1+z_{k+1} \bar z_{k+1}}-\frac
  { z_k} {1+z_k \bar z_k}\right)\frac{\bar z_{k+1}-\bar z_k}2\nonumber\\
  \HH_0(\bar z_1,z_1)=\alpha\frac{\bar z_1-1}{1+ z_1 \bar
  z_1}+\gamma \frac{z_1 (1-\bar z_1)}{1+z_1 \bar z_1};\quad \HH_L(\bar
  z_L,z_L)=\delta\frac{\bar z_L-1}{1+ z_L \bar z_L}+\beta
  \frac{z_L (1-\bar z_L)}{1+z_L \bar z_L}
\end{eqnarray}
Initial and final conditions on the fields $z,\bar z$ are imposed by
the delta functions in \eref{eq:propninf} and read
\begin{eqnarray} \label{eq:timeboundaryzz}
  \frac{\bar z_k^iz_k^i}{1+\bar z_k^iz_k^i} = \rho_k^i  \qquad
  \frac{\bar z_k^fz_k^f}{1+\bar z_k^fz_k^f} = \rho_k^f
\end{eqnarray}
\subsubsection{Density field.}
To get more insight on the physics of this field theory, we introduce a
new parametrization 
\begin{equation}
  \label{eqn:transfo}
  z_k = \frac{\rho_k}{1-\rho_k}\,e^{-\hat \rho_k} \:,\qquad \bar z_k =
  e^{\hat \rho_k}
\end{equation}
{so that
\begin{equation}
  \label{eq:corres}
  \rho_k= \frac{z_k \bar z_k}{1+z_k \bar z_k} = \frac 1 {2j}\langle
  z_k | j+S^z_k | z_k \rangle
\end{equation}
plays the role of a density. The transformation \eref{eqn:transfo} is
such that $\hat \rho$ is canonically conjugated to $\rho$.}  The time
boundary conditions \eref{eq:timeboundaryzz} on the field can be
written, as expected, as
\begin{equation}
  \label{eq:temporalBCrho}
  {\rho_k}(0)={ \rho_k^i} \:, \qquad {\rho_k}(T)={ \rho_k^f}
  \:,
\end{equation}
whereas $\hat \rho_k(0)$ and $\hat \rho_k(T)$ are unconstrained (for
details, see \ref{app:action_func}). {This highlights the
correspondence between the field ${\bol \rho}$ and the actual density
of the system, as do \eref{eq:corres}}. From \eref{eq:classicS} and
\eref{eqn:transfo}, one sees the correspondence between spins and densities
\begin{equation} \label{eq:classicSrho}
\langle z_k | S_k^+ | z_k\rangle = 2j(1-\rho_k)e^{\hat \rho_k} \:, \qquad
\langle z_k | S_k^- | z_k\rangle = 2j\rho_k\, e^{-\hat \rho_k} \:, \qquad
\langle z_k | S_k^z | z_k\rangle = j(2\rho_k -1)
\end{equation}
This yields for the Hamiltonian
\begin{eqnarray}
  \HH(\hat \rho,\rho) = \frac 1 2 \sum_{k=1}^{L-1}\left\{ (1-\rho_k)
  \rho_{k+1} \left[\rme^{\hat \rho_k - \hat \rho_{k+1}}-1\right]+ 
  \rho_k (1-\rho_{k+1})\left[\rme^{\hat \rho_{k+1}-\hat
      \rho_k}-1\right]\right\}\\
  +\left\{\alpha(1-\rho_1)(e^{\hat
    \rho_1}-1)+\gamma \rho_1(\rme^{-\hat\rho_1}-1)+\delta
  (1-\rho_L)(\rme^{\hat\rho_L}-1)+\beta \rho_L
  (\rme^{-\hat\rho_L}-1)\right\}\nonumber
\end{eqnarray}
Making the change of variables in the action  gives
\begin{eqnarray}
  P(\bol{ n_f},T;\bol{ n_i},0)&=\int \D\bol{ \hat \rho}\D \bol{
    \rho}\exp\{-S[\bol{ \hat \rho},{\bol \rho}]\}\\
  \label{eq:action1site}
  S[\bol{\hat\rho},\bol\rho]&= 2j \int_0^T \rmd t \left[ \sum_k
    \hat\rho_k \dot\rho_k - \HH(\hat \rho,\rho)\right]
\end{eqnarray}
where one integrates over fields $\bol\rho$ satisfying the temporal
boundary conditions \eref{eq:temporalBCrho}.
\subsubsection{Connection with Doi-Peliti variables}
In the previous subsections we took advantage of the SU(2) coherent
states to construct the path-integral representation of the evolution
operator. A more standard approach, based on Doi-Peliti formalism,
could have been followed. Starting from the density fields $\rho, \hat
\rho$, the usual bosonic coherent states can be obtained through a
Cole-Hopf transformation~\cite{Biroli2007}:
\begin{equation}
  \label{eqn:cole-hopf}
  \rho = \phi \phi^*;\quad \hat \rho = \log  \phi^*
\end{equation}
which  in the $z,\bar z$ variables reads
\begin{equation}
  \label{eqn:doipelitiSU2bis}
  \phi^* =  \bar z;\quad  \phi =\frac { z}{1+z \bar z}
\end{equation}
From \eref{eqn:hamiltonianz} one sees that the bulk Hamiltonian
$\HH_B$ reads in these variables
\begin{equation}
  \HH_B(\bol\phi,\bol \phi^*)=-\frac 1 2 \sum_k \phi_k \phi_{k+1}
  (\phi^*_{k+1}-\phi^*_k)^2+(\phi_{k+1}-\phi_k)(\phi^*_{k+1}-\phi^*_k)
\end{equation}

To understand  the meaning of the $\phi, \phi^*$ variables, we recall the
action of the spin operators in the coherent state representation:
\begin{eqnarray}
  \langle z_k|S^+_k|\psi\rangle &=& \left[2j \bar z_k - \bar z_k^2
    \frac\partial{\partial \bar z_k}\right] \langle z_k | \psi \rangle
  \nonumber\\ \langle z_k|S^-_k|\psi\rangle &=& \frac\partial
              {\partial \bar z_k} \langle z_k|\psi\rangle \nonumber
              \\ \langle z_k|S^z_k|\psi\rangle &=& \left[\bar z_k
                \frac{\partial}{\partial \bar z_k}-j \right] \langle
              z_k|\psi\rangle\label{eqn:SkAction}
\end{eqnarray}
The matrix element used to construct the path integral
(\ref{eq:classicS}) are  given in terms of $\phi,\phi^*$ by
\begin{equation}
  \label{eqn:spinottt}
  \langle z_k | S_k^+ | z_k\rangle= 2j \phi^*_k -(\phi^*_k)^2 (2j\phi_k); 
  \qquad 
  \langle z_k | S_k^- | z_k\rangle =2j \phi_k;
  \qquad 
  \langle z_k | S_k^z | z_k\rangle = \phi^*_k (2 j \phi_k)-j \;,
\end{equation}
and one sees by comparing \eref{eqn:SkAction} and \eref{eqn:spinottt} that
\begin{equation}
  \label{eqn:doipelitiSU2}
  \phi^* = \bar z;\qquad \phi = \frac 1 {2j} \frac{\partial}{\partial
    \bar z}
\end{equation}
$\phi$ and $\phi^*$ satisfy the usual bosonic commutation relation.
They correspond to the usual Doi-Peliti~\cite{Doi1976,Peliti1985}
representation of bosons. We thus see that one can construct a path
integral in terms of the variables $\phi$ and $\phi^*$ either by a
Cole-Hopf transformation, or by directly expressing the spin operators
in the representation (\ref{eqn:SkAction}) and then constructing the
path integral treating $ \bar z$ and $\frac 1 {2j}
\frac{\partial}{\partial \bar z}$ as conjugate bosons~\footnote{One
  must however proceed with care as states with more than $2j$
  particles are not physical and are difficult to handle using bosonic
  coherent state. See for instance~\cite{fvw2001}.}.  In what follows,
however, it will be useful not to lose sight of the SU(2) symmetry.

\if{
  The spin operators act on the coherent states representation as
  \begin{eqnarray}
    \langle\psi|S^-_k|z_k\rangle &=& \left[-z_k^2
      \frac{\partial}{\partial z_k}+ 2jz_k \right]
    \langle\psi|z_k\rangle \nonumber \\ \langle\psi|S^+_k|z_k\rangle
    &=& \frac{\partial}{\partial z_k} \langle\psi|z_k\rangle \nonumber
    \\ \langle\psi|S^z_k|z_k\rangle &=& \left[z_k
      \frac{\partial}{\partial z_k}-j \right] \langle \psi|z_k\rangle
  \end{eqnarray}
  which means that
  \begin{equation}
    S^-_k \to -z_k^2 \frac{\partial}{\partial z_k}+ 2jz_k;\qquad S^+_k
    \to \frac{\partial}{\partial z_k};\qquad S^z_k \to z_k
    \frac{\partial}{\partial z_k}-j
  \end{equation}
  This leads us to introduce the coordinate-momenta (bosonic) like variables
  \begin{equation}
    a_k=\frac{1+z_k}2,\qquad a_k^\dagger=\frac 1 j \frac{\partial}{\partial z_K}
  \end{equation}
  which satisfy the usual commutation relation
  \begin{equation}
    [a^\dagger_k,a_{k'}]=\frac{\delta_{k,k'}}{2j}
  \end{equation}
  yielding
  \begin{equation}
    S^+_k=a_k^\dagger,\qquad S^-_k=2(2a_k-1)-(2a_k-1)^2a_k^\dagger,\qquad
    S^z_k=(2a_k-1) a_k^\dagger-1
  \end{equation}
  
  In what follows we have found extremely useful to follow the same
  procedure with a new set of axes, obtained after a rotation of
  angle $\pi/2$ around the $y$-axis:
  \begin{equation}
    {S'}^x_k={S}^z_k\qquad {S'}^z_k=-S^x_k\qquad {S'}^y_k=S^y_k
  \end{equation}
  It leads to the corresponding 'bosonic' variables
  \begin{equation}
    {S'}^+_k=\bar F_k\qquad {S'}^-_k=2(2F_k-1)-(2F_k-1)^2 \bar F_k\quad
    {S'}^z_k=(2F_k-1)\bar F_k-1
  \end{equation}
  with the commutation relation
  \begin{equation}
    [\bar F_k,F_{k'}]=\frac{\delta_{k,k'}}{2j}
  \end{equation}
}\fi

\subsubsection{Hydrodynamic limit}
So far, the action is still a discrete sum over the whole lattice. We
shall now take the full hydrodynamic limit to describe the evolution
of smooth profiles on diffusive timescales. We thus introduce a space
parametrization $x_k=\frac{k}L$ and rescale the time $t \to L^2 t$. At
the macroscopic level, the density profiles are smooth functions and
discrete gradients can be replaced by continuous ones:
\begin{equation}
  \rho_{k+1}-\rho_k \to \frac {\grad \rho(x_k)}L,\quad
  \hat\rho_{k+1}-\hat\rho_k \to \frac {\grad \hat\rho(x_k)}L, \quad
  \frac 1 L\sum_{k=1}^{L-1}\to \int_0^1 \rmd x 
\end{equation}
For a  more rigorous approach, see~\cite{Spohn1983,Bertini2002}. The
first order in a $\frac 1 L$ expansion of the action then reads
\begin{eqnarray}
  \label{eqn:actiondyn}
  P(\bol{\rho^f},t_f;\bol{\rho^i},0)=\int\D {\hat \rho} \D \rho
  \,\rme^{-2j\,L\, S[\rho,\hat\rho]}\\ S[\rho,\hat\rho]=\int_0^{t_f}
  \int_0^1 \rmd x \rmd t \big\{ \hat \rho \partial_t
  \rho-\HH[\rho,\hat\rho]\big\};\qquad \HH[\rho,\hat\rho]= \frac 1
  2 \sigma (\grad \hat \rho)^2 - \frac 1 2 \grad \rho \grad \hat \rho
  \label{eqn:actiondynham}
\end{eqnarray}
where {$\sigma=\rho(1-\rho)$}.

In  \ref{app:hydrolim} we show that 
   fields $\rho$ and $\hat \rho$
 are constrained, in the hydrodynamic limit, to satisfy
 the  spatio-temporal boundary conditions
\begin{eqnarray}
  \forall t,\quad \rho(0,t)=\rho_0=\frac{\alpha}{\alpha+\gamma},\quad
  \rho(1,t)=\rho_1=\frac{\delta}{\delta+\beta},\quad\hat\rho(0)=
\hat\rho(L)=0\label{eqn:BC1}\\ \rho(x,0)=\rho_i(x),\quad
  \rho(x,T)=\rho_f(x) \label{eqn:BC2}
\end{eqnarray}
If
$\alpha,\beta,\gamma,\delta\sim{\cal O}(1)$ at the microscopic level,
these conditions are strict, in the sense that the fields do not
fluctuate in the borders~\footnote{{The probability
  to observe a smooth profile such that $\rho(0)\neq \rho_0$ scales as
  $e^{-jL^2}$.}}, {whereas $\alpha,\beta,\gamma,\delta\sim {\cal
    O}(L^{-1})$ would also allow fluctuations of $\rho$ and $\hat \rho$ at
  the boundaries.}

Due to the correspondence
\eref{eqn:hamspindiscret},\eref{eq:classicSrho} 
with the spin operators, the whole process
described here simply amounts to taking the classical limit of a
Heisenberg spin chain with some particular boundary conditions. One
can indeed check that the Hamiltonian \eref{eqn:actiondynham}
corresponds to
\begin{equation}
  \label{eqn:hamiltonianspin}
  \HH_B= -\frac 1 2 \int \text{d} x  \grad \bol S \cdot \grad \bol S
\end{equation}
{where the classical spins~\cite{foot4a} are defined as
  $S^{x,y,z}=\lim_{j\to\infty} (2j)^{-1} \langle
  z|S^{x,y,z}|z\rangle$}. The spatial boundary conditions in terms of
classical spins are then given by
\begin{eqnarray}
  \label{eqn:spinBC}
  S^z(0)&=\rho_0-\frac 12 \qquad\qquad\qquad &S^z(1)=\rho_1-\frac 12\\
  S^x(0)&=\frac 12 \qquad\qquad\qquad &S^x(1)=\frac 12\nonumber\\
  S^y(0)&=\frac{1-2\rho_0}{2 \rmi} \qquad\qquad &S^y(1)=\frac{1-2\rho_1}{2\rmi}\nonumber\\
  S^+(0)&=1-\rho_0 \qquad\qquad &S^+(1)=1-\rho_1\nonumber\\
  S^-(0)&=\rho_0 \qquad\qquad &S^-(1)=\rho_1\nonumber
\end{eqnarray}

\pagebreak

\section{Fluctuating hydrodynamics}
\label{sec:FH}
Let us briefly review in this section the construction of the action
starting from fluctuating hydrodynamics and using the
Martin-Siggia-Rose~\cite{MSR1973},
DeDominicis-Janssen~\cite{DeDominicis1975,Janssen1976}
formalism. 

In terms of the instantaneous current $J(x)$ at site $x$, defined by
the continuity equation, {the evolution of the density is given by
  the stochastic equation}
\begin{equation}
  \label{stochdyn}
  \dot \rho = -\nabla J; \qquad \;\;\,J = - \frac{1}{2} \nabla
  \rho - \sqrt{\sigma} \eta;\qquad {\rho(0)=\rho_0;\qquad \rho(1)=\rho_1}
\end{equation}
where $\eta$ is a white noise of variance $1/(2j\,L)$ and
$\sigma=\rho (1-\rho)$. This is the usual formula for the
fluctuating hydrodynamics of the exclusion process~\cite{Spohn1983}.
We write this as a sum over paths and noise realizations with a delta
function imposing the equations of motion {and a Gaussian weight
  for the noise}:
\begin{equation}
  P({\bol \rho^f},t^f;{\bol \rho^i},0)=\int\D {\eta} \D \rho \,\;
 \delta[ \dot \rho + \nabla J] \;\;
\rme^{-2j\,L\, \int_0^{t_f}  \int_0^1 \rmd x \rmd t \; \frac{1}{2} \eta^2(x,t)}
 \end{equation}
Exponentiating the functional delta function with the aid of a
function $\hat\rho(x,t)$:
\begin{equation} 
  P({\bol \rho^f},t_f;{\bol \rho^i},0)   =\int \D \eta \,\D \rho \, \D \hat\rho \;
\rme^{ -2jL \{\int_0^{t_f} \int_0^1 \rmd x \rmd t \; \hat\rho ( \dot \rho + \nabla
J)+ \frac{1}{2} \eta^2(x,t) \} }
 \end{equation}
where $\hat \rho$ is integrated along the imaginary axis. Integrating
by parts, we obtain:
\begin{equation} 
  P({\bol \rho^f},t_f;{\bol \rho^i},0)   =\int \D \eta\, \D \rho \, \D \hat\rho \;
\rme^{-2jL \{\int_0^{t_f} \int_0^1 \rmd x \rmd t \; (\hat\rho  \dot \rho + \nabla \hat\rho
(\frac{1}{2} \nabla
  \rho +  \sqrt{\sigma} \eta))+ \frac{1}{2}\eta^2(x,t) \} }
 \end{equation}
We can now integrate away the noise:
\begin{equation} 
\label{eqnsldfhk}
  P({\bol \rho^f},t_f;{\bol \rho^i},0)  =\int \D \rho \, \D \hat\rho \; \rme^{-2jL \{\int_0^{t_f} \int_0^1 \rmd
    x \rmd t \; (\hat\rho \dot \rho +\frac{1}{2} \nabla \hat\rho
    \nabla \rho - \frac{1}{2} (\nabla \hat\rho)^2 \sigma \} }
 \end{equation}
which reads
\begin{equation} 
  P({\bol \rho^f},t_f;{\bol \rho^i},0) =\int \D \rho \, \D \hat\rho
\; \rme^{-2jL \{\int_0^{t_f} \int_0^1 \rmd x \rmd t \; (\hat\rho \dot \rho
- {\cal{ H}}) \} }
  \label{eqn:flucthydro}
 \end{equation}
to obtain
\begin{equation} 
 {\cal{ H}}  = \frac{1}{2} \left[\sigma (\nabla \hat\rho)^2-  \grad \hat\rho
\nabla \rho\right]
 \end{equation}
which is {equivalent to} (\ref{eqn:actiondyn}). The
paths are constrained to be $\rho_i(x)$ and $\rho_f(x)$ at initial and
final times, respectively. The values of $\hat \rho$ are
unconstrained, which is in agreement with the fact that this is a
Hamiltonian problem with two sets of boundary conditions.
The construction above can thus be seen as a formal
Hubbard-Stratonovich transformation to introduce the $\hat\rho$
field. 

Let us finally note that from equation \eref{stochdyn} and \eref{eqnsldfhk}, one
sees that
\begin{equation} 
  P({\bol \rho^f},t_f;{\bol \rho^i},0)  =\int \D \rho \, \D \hat\rho \; \rme^{-2jL \{\int_0^{t_f} \int_0^1 \rmd
    x \rmd t \; ( J\grad \hat\rho  +\frac{1}{2} \nabla \hat\rho
    \nabla \rho - \frac{1}{2} (\nabla \hat\rho)^2 \sigma \} }
 \end{equation}
Formally integrating over $\grad \hat\rho$ gives back the usual
fluctuating hydrodynamics~\cite{Bertini2001,Jordan2004,Pilgram2003}
\begin{equation} 
  P({\bol \rho^f},t_f;{\bol \rho^i},0)  =\int \D \rho \; \exp\Big[-2jL\int_0^{t_f} \int_0^1 \rmd
    x \rmd t \; \frac{(J  +\nabla \rho/2)^2}{2 \sigma}  \Big]
 \end{equation}

In all this we have been very sloppy about the spatial conditions
$\hat\rho$ should satisfy: see \ref{app:BC} for details.

\pagebreak

\section{Large deviations in the coarse-grained limit}
\label{sec:LDLNL}

In this section we review the steps leading from the fluctuating
theory to a non-fluctuating Hamiltonian dynamics, valid in the low-noise
limit --- itself arising in the large coarse-graining limit. 

\subsection{Classical solutions}

In order to calculate a probability we have to evaluate
\begin{equation}
\label{eqn:probaprof}
  P(\bol{ \rho^f},T)= \int \rmd {\bol{ \rho^i}} P(\bol{
    \rho^f},T;\bol{ \rho^i},0) P(\bol{ \rho^i},0)
\end{equation}
which is true for any time $T$. Using the path integral expressions of
the previous sections (cfr (\ref{eq:propninf})), one then gets the
sum of the exponential of the action
\begin{equation}
  \label{eqn:probarhostar}
  P(\bol{ \rho^f},T)= \int \rmd {\bol{ \rho^i}} \int {\cal
    D}[\rho,\hat \rho] \rme^{-2j\,L\, S[\rho,\hat\rho]} P(\bol{
    \rho^i},0)
\end{equation}
over trajectories with initial and final profiles $\bol{ \rho^i}$ and
$\bol{ \rho^f}$.
To leading order in $jL$, we have that $ P(\bol{ \rho^f},T;\bol{
  \rho^i},0) $ is dominated by the trajectories extremalizing the action
\eref{eqn:actiondyn}
\begin{equation}
  S= \int dx dt \; [ \hat\rho \dot \rho - {\HH} ]
\end{equation}
i.e. satisfying Hamilton's equations:
\begin{eqnarray}
  {\dot \rho}(x,t) = \frac{\delta }{\delta \hat {\rho}(x,t)} \int dx'
  dt' \HH[\rho(x',t'),\hat\rho(x',t')]\\
  {\dot {\hat \rho}}(x,t) = -\frac{\delta }{\delta {\rho}(x,t)} \int dx'
  dt' \HH[\rho(x',t'),\hat\rho(x',t')]
\end{eqnarray}
These are completely determined (at least up to a discrete set of
trajectories) by the initial and final values of $\rho(x)$.  What we
have outlined is the exact analogue of the way classical trajectories
dominate the path-integral in the semi-classical limit $\hbar\to 0$ in
quantum mechanics. Similar approaches have been used many times, as
for instance to analyze the noisy Burgers equation~\cite{Fogedby1999}
or in reaction diffusion systems~\cite{Kamenev2004}.

In the case of the SSEP, Hamilton's equations read:
\begin{eqnarray}
\label{classicalequations1}
  \dot \rho(x) = \frac 1 2
  \Delta \rho - \grad [\sigma\grad \hat\rho]\\ \dot {\hat\rho}(x)
  =(\rho-\frac 1 2) (\grad \hat\rho)^2 -
  \frac 1 2 \Delta\hat\rho
\end{eqnarray}
 \eref{classicalequations1} is a conservation equation
and thus defines a current through $\dot \rho = -\grad J_\rho$ where
\begin{equation}
  \label{eqn:currentrhohatrho}
  J_\rho[\rho,\hat \rho]=-\frac 1 2 \grad \rho + \sigma \grad \hat \rho
\end{equation}
To determine the probability of a
transition between a profile $\bol{ \rho^i}$ and another profile
$\bol{ \rho^f}$ in a time $T$, one thus has to find the trajectory
$\big(\rho(x,t),\hat\rho(x,t)\big)$ that solves
\eref{classicalequations1} and satisfies the appropriate 
boundary conditions.  The action of this trajectory then yields the
logarithm of the probability of the transition.

We are thus led to solving a classical Hamiltonian {\em field}
problem, where initial and final positions are fixed and momenta
unknown. This is in general very difficult -- even numerically, where
one has to solve a `shooting' problem to reach the desired final
configuration at the right time~\cite{foot5}.

For completeness, let us write the equations of motion in terms of classical
spins satisfying the Poisson bracket algebra~\cite{foot4}:
\begin{equation}
\{S^z(x),S^\pm(y)\}=\pm S^\pm(x) \; \delta(x-y)  \;\;\;\; ; \;\;\;\; 
\{S^+(x),S^-(y)\}=2S^z \; \delta(x-y)
\end{equation}
which corresponds to the quantum commutations~\eref{eq:CommutQuantumSpins}
and yields the equation of motion in terms of spin variables through
\begin{equation}
  \label{hamspin}
  \dot {\bol S} = \{ {\bol S}, H\}= 
 \rmi  \bol {\Delta S} \wedge \bol S
\end{equation}

\subsection{Downhill trajectories}

For classical equations deriving from a stochastic problem, there
always exists a class of trajectories which are easy to find: those
that are overwhelmingly the most likely {in the low noise limit}. Here
and in what follows, we shall call these `downhill' trajectories.  If
one remembers that the dynamical action corresponds to the stochastic
equation \eref{stochdyn}, one obtains these solutions by directly
putting $\eta=0$.  They correspond to the following solutions of the
system (\ref{classicalequations1}):
\begin{equation}
\label{downequations1}
\dot \rho(x) = \frac 1 2  \Delta \rho \qquad {\hat\rho}(x,t)=0
\end{equation}
The solution corresponds to diffusive relaxation towards the linear
stationary profile $\bar \rho(x)=(1-x)\rho_0+x\rho_1$.
 The corresponding action is zero,
in agreement with the fact that the corresponding probability is 1,
which simply means that  an initial
configuration $\rho(x)$ almost surely relaxes towards the
stationary state. 
The stationary profile $\hat \rho=0, \rho=\bar\rho$
is  a {\em hyperbolic} fixed point of the dynamics in the full
phase-space, since it is missed 
as soon as $\hat \rho \neq 0$ in the initial condition.

\subsection{Large deviation function from extremal trajectories}
\label{sec:LDFfromACtion}

As shown by equation \eref{eqn:probaprof}, the probability $P(\rho^*)
\sim e^{-N{\cal{F(\rho^*)}}}$ to observe a profile $\rho^*$ is the
average probability of going from an initial profile $\bol \rho_i$ to the
profile $\rho^*$. The logarithm of this transition probability is
given in the large $N$ limit by the `classical' action of a trajectory
starting in $\bol{ \rho^i}$ and arriving in a time $T$ in the
configuration $\bol{ \rho^f}=\rho^*$, and satisfying
(\ref{classicalequations1}).

\begin{figure}
  \begin{center}
    \includegraphics[width=10cm,angle=-90]{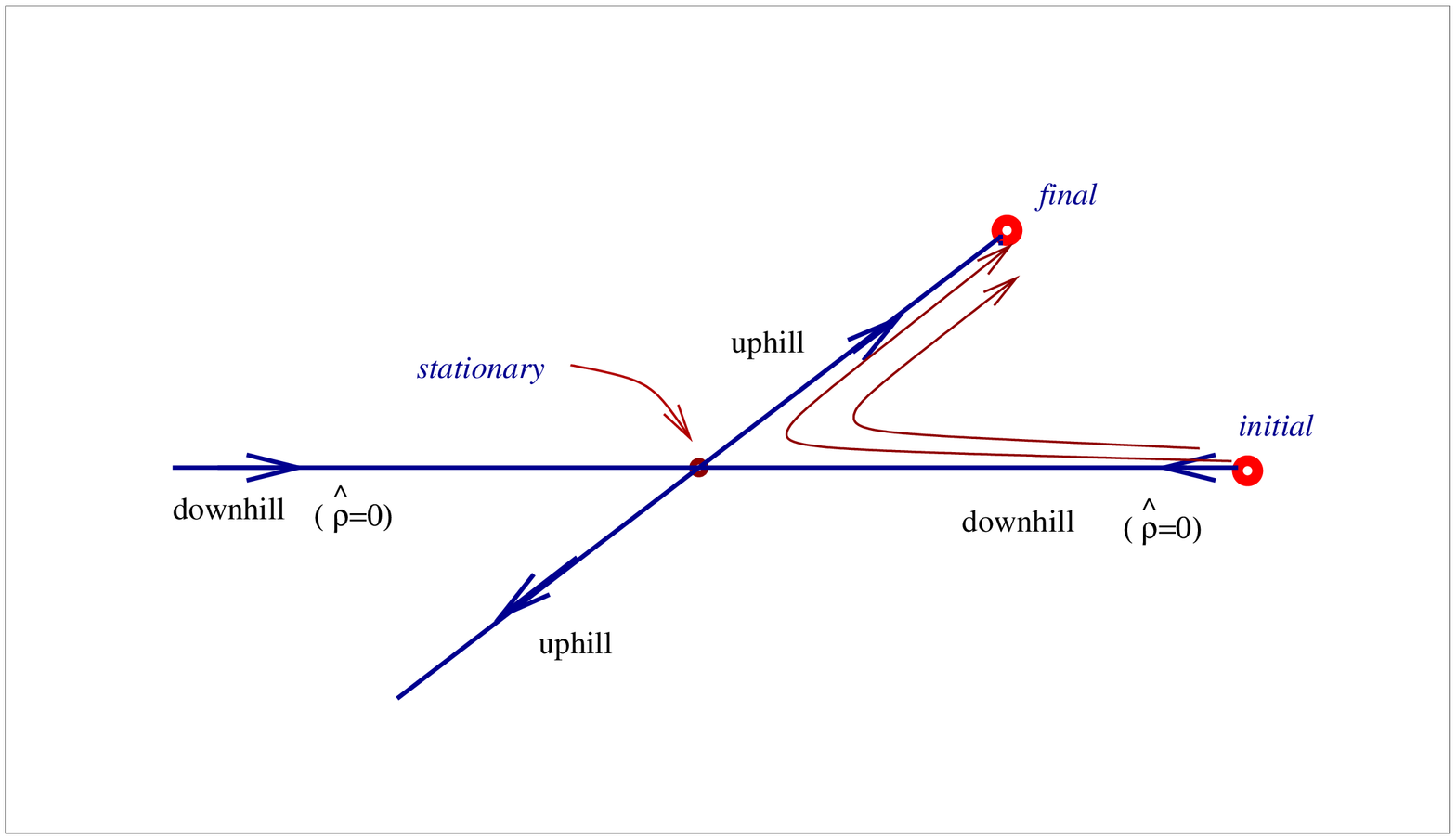}
    \caption{ `Downhill' (noiseless) trajectories have $\hat
      \rho(t)=0$. Trajectories reaching an arbitrary point at long
      times are ``near-misses'' from the stationary point, and can be
      decomposed in {the long-time} limit in a `downhill',
      followed by an `uphill' trajectory flowing into, and out of the
      stationary point, respectively.}
    \label{fig:extremetraj}
  \end{center}
\end{figure}

How can a trajectory just reach a generic configuration $\rho^*$ at a
very large time $T \to \infty$? The only possibility is that it takes
a hyperbolic trajectory that falls in a finite time in the vicinity of
the stationary profile, stays there almost all the time, and then goes
to the profile $\rho^*$ in a finite time. This is illustrated in
figure \ref{fig:extremetraj}: trajectories that matter at long times
are thus near-misses of the stationary points. The first part of such
a trajectory (essentially the diffusive relaxation towards the
stationary profile) has almost zero action.  Thus $P(\bol{ \rho^i},0
\to \rho^*,T)=P(\bar\rho,\rho^*,T')$ where $T'$ is a time which
differs from $T$ by a finite contribution, not relevant in the
$T\to\infty$ limit: the transition probability $P(\bol{ \rho^i},0 \to
\rho^*,T)$ is independent of $\rho^i$ in the long time limit and
equation \eref{eqn:probaprof} becomes
\begin{equation}
  \label{eqn:probaprof2} P(\bol{ \rho^*},T\to\infty)= \int \rmd {\bol{
      \rho^i}} P(\bol{ \rho^i},0) P(\bar\rho,\rho^*,T'\to\infty) =
  P(\bar\rho,\rho^*,T'\to\infty)
\end{equation}

To determine the probability to observe a given profile $\rho^*$, one
thus has to find the extremal trajectory which starts at $t=-\infty$
in the stationary profile and arrives at $t=0$ in the desired profile
$\rho^*$. Its action then yields the large deviation function
${\cal{F}[\rho^*]}$
\begin{equation}
  P(\rho^*)=  P(\bar\rho,t=-\infty,\rho^*,t=0)
\end{equation}
We have assumed here that there are no
metastable states. In cases in which many such states exist, one has to
consider trajectories falling into each one of them, and also making 
jumps between them before reaching the final point.

\pagebreak

\section{Detailed balance relation, relaxation-excursion symmetry}
\label{sec:DBR}

Finding the trajectory that reaches a given density profile starting
from the vicinity of the stationary one is in general a difficult
task.  There is however a class of systems for which this calculation
greatly simplifies: those for which there is a {\em detailed balance}
symmetry -- playing the role of a time-reversal -- that relates the
paths followed by the system in a rare (noise-induced) excursion with
the relaxation back to equilibrium.  This {\em Onsager-Machlup}
symmetry allows one to compute the rare excursions (in general
difficult) from the sole knowledge of the relaxations (easy, as
explained in the previous section). This is for instance the case for
a SSEP in contact with reservoirs of equal densities $\rho_0=\rho_1$,
as we shall see below.

At the level of operators, detailed balance simply says that the
evolution operator and its adjoint are related by a similarity
transformation. Here, we are only interested in symmetries at the
level of the action, which can be read as a canonical transformation
followed by time reversal, leaving the action invariant up to boundary
terms (See \ref{sec:symmetryFromMicrotoMacro}).

To make this explicit in our case, we write the Hamiltonian density as
\begin{equation}
  \HH[\rho,\hat\rho]= \frac 1 2 \grad \hat \rho \,\sigma \grad \left(
  \hat\rho -\log \frac \rho{1-\rho} \right)= \frac 1 2 \grad \hat \rho
\,  \sigma \grad\left( \hat\rho -\frac{\delta V_\rho}{\delta \rho}
  \right)
\end{equation}
{where $V_\rho$ is the equilibrium entropy:}
\begin{equation}
  V_\rho=\int \rmd x [\rho \log\rho + (1-\rho)\log(1-\rho)]
\end{equation}
The transformation we are looking for is given in two steps
\cite{Janssen1979}: {\em i)}
{the canonical transformation}
\begin{equation}
  \label{DB1}
  \hat \rho \to \hat \rho + \frac{\delta V_\rho}{\delta \rho} \;\;\;\; ;
 \;\;\;\; \rho \to \rho
\end{equation}
followed by: {\em ii)} {a time reversal}
\begin{equation}
\label{DB2}
  (\hat\rho, \rho, t) \to (-\hat \rho, \rho, T-t)
\end{equation}
In terms of the current \eref{eqn:currentrhohatrho}, this reads:
\begin{equation}
  [\rho(x,t),J(x,t)] \;\;\;  \rightarrow \;\;\;  [\rho(x,-t),-J(x,-t)]
  \label{ddbb}
\end{equation}
the  meaning of which is transparent.
The new, `time-reversed' variables are:
\begin{equation}
  \label{eqn:transfoDBrhos}
\rho_{\mbox{\tiny TR}}(x,t)=\rho(x,-t) \;\;\; ; \;\;\; \hat
\rho_{\mbox{\tiny TR}}(x,t)=- \hat \rho(x,-t) + \frac{\delta
    V_\rho}{\delta \rho} (x,-t)
\end{equation}

It is easy to check that
  (\ref{DB1},\ref{DB2}) map the action into
\begin{equation}
  \label{eqn:changeDB}
  S[\hat \rho, \rho]  \to [V_\rho]_0^T +  \int \text{d}
  t \text{d}x \left\{ \hat \rho_{\mbox{\tiny TR}} \dot \rho_{\mbox{\tiny TR}}
 -  {\cal H}[\hat \rho_{\mbox{\tiny TR}},\rho_{\mbox{\tiny TR}}]\right\}
\end{equation}
The space and time  boundary conditions are transformed from \eref{eqn:BC1} and
\eref{eqn:BC2} to
\begin{eqnarray}
  \rho_{\mbox{\tiny TR}}(x,0)=\rho^*(x)\qquad
  \rho_{\mbox{\tiny TR}}(x,T)=\bar\rho(x)\\ \rho_{\mbox{\tiny TR}}(0,t)=
\rho_0\qquad \rho_{\mbox{\tiny TR}}(1,t)=
  \rho_1\qquad\hat \rho_{\mbox{\tiny TR}}(0,t)
= \log\frac{\rho_0}{1-\rho_0}\qquad\hat
  \rho_{\mbox{\tiny TR}}(1,t)= 
\log\frac{\rho_1}{1-\rho_1}\label{eqn:spatialBCDB}
\end{eqnarray}
The problem is thus recast into finding a trajectory starting in
$\rho^*$ and relaxing towards the stationary profile. When the system
is at equilibrium, in contact with two reservoirs imposing the same
density on the two boundaries ($\rho_0=\rho_1$), the zero-noise
diffusive trajectory
\begin{equation}
  \label{eqn:DBSolution}
  \hat\rho_{\mbox{\tiny TR}}(x,t)= \text{C}^{st}=
\log\frac{\rho_0}{1-\rho_0}\qquad \dot \rho_{\mbox{\tiny TR}}=\frac 1 2 \Delta \rho_{\mbox{\tiny TR}}
\end{equation}
is a legitimate solution of the classical equations which satisfies
the new boundary conditions in space and time. The action of such a
trajectory is
\begin{equation}
{\int \text{d} t \text{d}x \left\{ \hat \rho_{\mbox{\tiny TR}} \dot \rho_{\mbox{\tiny TR}} - {\cal
  H}\right\}}= \log\frac{\rho_0}{1-\rho_0} \int\rmd x (\rho_0-\rho^*)
\end{equation}
Together with the boundary terms of \eref{eqn:changeDB}, one gets for
the large deviation function, {\em in terms of the original variables}
\begin{equation}
  S[\rho^*]= \int\rmd x \left[(1-\rho)\log\frac
  {1-\rho}{1-\rho_0}+\rho\log\frac\rho{\rho_0}\right]
\end{equation}
which is the usual equilibrium entropy~\cite{Bertini2002}.
The computation of the extremal trajectory going from $\bar \rho$ to
$\rho^*$ has thus been made possible by the time-reversal
connection between excursion and relaxation induced by the detailed
balance relation. Thanks to the mapping \eref{DB1} and \eref{DB2}, one
just has to find a trajectory going from $\rho^*$ to $\bar \rho$ --
 a relaxation -- and from it obtain an excursion. 


\subsection{System driven out of equilibrium - 
Violation of detailed balance and loss of Onsager-Machlup symmetry}

Let us now consider why this `trick' does not work when the system is
driven out-of-equilibrium by the boundaries. 
  One can still make the
transformation \eref{DB1} and \eref{DB2}, as the bulk dynamics 
satisfies detailed balance. {\em However,
if $\rho_1\neq\rho_0$, \eref{eqn:DBSolution} is not acceptable as
$\hat \rho=\text{C}^{st}$ does not satisfy the spatial boundary
conditions}. We conclude that
 the most probable excursion from $\bar\rho$ to $\rho^*$
is consequently  not the time reversed of a diffusive
relaxation: the Onsager-Machlup symmetry is
  broken by the boundaries.

\pagebreak

\section{Driven exclusion process: remarkable changes of variables}

\label{sec:FetHatF}

As we have seen,
in out of equilibrium systems  detailed balance is violated -- for
example by the boundary conditions -- and there is no
{simple} relation between excursions and relaxations. One thus
cannot, in general, compute easily the rare excursions away from the
stationary state, and from it the large deviation function. 
Even if we have been able to reduce the computation of
large-deviation functions to the solution of a problem of classical 
dynamics, we are not able to solve for its trajectories in a closed
way. It may then come as a surprise that BDGJL~\cite{Bertini2001,Bertini2002} were able to uncover what
amounts to a series of changes of variables, which end up by mapping
the driven problem into one where there is a simple time-reversal symmetry
between excursions and relaxations. 
This allowed them to compute the excursions in terms of the
relaxations in the new variables, and then work their way back to the
original variables {\em in which the symmetry does not hold}.

In this section we shall paraphrase their derivation, to emphasize how
surprising it is. In the next section we shall argue that at the
bottom of this is the (also very surprising) fact that the SSEP driven
out of equilibrium by the boundaries can be mapped back through a
change of variables, in the hydrodynamic limit, to an equilibrium
SSEP.

Let us first note that the choice of axes we have used to write the
hydrodynamic limit Hamiltonian \eref{eqn:hamiltonianspin}
 in terms of spin variables
$ \HH_B=-\frac 1 2 \int \rmd x {
  \bol \grad S} \cdot { \bol \grad S}$ is
arbitrary. 
Only the boundary conditions (\ref{eqn:spinBC}) break the rotation invariance.


Let us make a transformation:
\begin{equation}
  S_x'=S_z\qquad S_z'=-S_x\qquad S_y'=-S_y \; ;
\end{equation}
 a rotation of angle $\pi/2$ around the $y$ axis followed by a
reflexion with respect to the x'-z' plane. 
In terms of the coherent-state coordinates -- the stereographic
representation of classical spins --- 
this transformation is given by the simple homography
\begin{equation}
  \bar z'=\frac{z-1}{z+1}\qquad z'=\frac{\bar z-1}{\bar z+1}
\end{equation}
In this set of variables, the action reads after a lengthy but
straightforward computation
\begin{equation}
  S=\int \rmd x \left[\rho \log
    \frac{\rho}{\frac{1+z'}{2}}+
(1-\rho)\log\frac{(1-\rho)}{\frac{1-z'}2}\right]+S[z',\bar
    z']
\end{equation}
As this change of variable corresponds to a symmetry of the
Hamiltonian, the action $S[z',\bar z']$ is the same as the original
one where $z$ and $\bar z$ have been replaced by $z'$ and $\bar
z'$. It is thus obtained by taking the continuum limit of
\eref{eq:propninf}:
\begin{equation}
  S[z',\bar z']=\int dx dt\left\{\frac{\dot {\bar z}' z'}{1+z' \bar
  z'}+\frac 1 2 (\grad \bar z')^2 \left(\frac {z'} {1+z' \bar z'}\right)^2+\frac 1 2 \grad \bar z \grad \left(\frac {z'} {1+z' \bar z'}\right)\right\}
\end{equation}
{\em Of course, we could have arrived at the same point 
by defining at the outset the coherent
states in terms of the rotated operators.}

As previously, we can make a Cole-Hopf transformation to 
introduce Doi-Peliti like variables:
\begin{equation}
  \phi'=\frac {z'}{1+z'\bar z'}\qquad {\phi^*}'=\bar z'
\end{equation}
Surprisingly, as we  shall show below, this change of variables greatly
simplifies the problem. To stay as close as possible to the solution
introduced by BDGJL, we rather introduce a slightly different
set of variables
\begin{equation}
  \label{eqn:vraiableF}
  F=\frac{1+{\phi^*}'}2,\qquad \hat F= 2 \phi'
\end{equation}
with which the same simplification occurs.
In the $F,\hat F$ variables, the
action is given by
\begin{eqnarray}
  \label{eqn:actionF}
  S&= \int \rmd x \left[\rho \log
    \frac{\rho}{F}+(1-\rho)\log\frac{(1-\rho)}{1-F}\right]+S_F[F,\hat
    F] \\ S_F[F,\hat F]&= \int \rmd t \rmd x \left[\hat F \dot F +
    \frac 1 2 \hat F^2 \grad F^2 + \frac 1 2 \grad F \grad \hat
    F\right]
\end{eqnarray}
The overall change of variables from  $\rho, \hat \rho$ to $F, \hat F$
reads
\begin{equation}
  \label{eqn:rhotoF}
  F=\frac{\rho}{\rho + (1-\rho) e^{\hat\rho}};\qquad \hat F = (1-\rho)
  (e^{\hat \rho}-1) - \rho (e^{-\hat\rho}-1)
\end{equation}
In terms of these, the spatial boundary conditions read
\begin{equation}
  \label{eqn:boundariesF}
  F(0)=\rho_0\qquad F(1)=\rho_1\qquad \hat F(0)=0\qquad \hat F(1)=0
\end{equation}
while the equations of motion become
\begin{equation}
  \label{eqn:motionF}
  \dot F = \frac 1 2 \Delta F - \hat F (\nabla F)^2;\quad \dot {\hat F}
=-\frac 1 2 \Delta \hat F - \nabla \left[ {\hat F}^2 \nabla F \right]
\end{equation}

In particular, by analogy with the `downhill' zero-noise solutions
of the previous sections, we can try $\hat F(t)=0$, which corresponds to
\begin{equation}
  \dot F = \frac 1 2 \Delta F
\end{equation}
In the original variables, this solution reads 
\begin{equation}
  \hat\rho =0;\qquad \dot 
 \rho=\frac 1 2 \Delta \rho
\end{equation}
which is nothing but the diffusive relaxation to the linearly
stationary profile $\bar \rho$. Diffusion in $F,\hat F$ variables thus
corresponds to relaxation in $\rho, \hat \rho$.

\subsection{A second detailed-balance like symmetry}

Remarkably, the action \eref{eqn:actionF} has a detailed balance-like
symmetry, which is unrelated to the original physical one.  To see
this, integrate the last term of $S_F$ by parts~\cite{foot42}, so that
the action becomes:
\begin{equation}
  \label{eqn:actionFF}
  S_F[F,\hat F]= \int \rmd t \rmd x \left\{\hat F \dot F + \frac 1 2
     \hat F \grad F^2 \left[\hat F - \frac{ \Delta F}{\grad
     F^2}\right]\right\}
\end{equation}
which can also be written
\begin{equation}
   S_F[F,\hat F]= \int \rmd t \rmd x \left[\hat F \dot F + \frac 1 2
     \hat F \grad F^2 \left[\hat F - \frac{ \delta V_F}{\delta
         F}\right]\right\}\; ;
\end{equation}
where we have introduced the potential
\begin{equation}
  V_F=\int \rmd x \log \grad F
\end{equation}
The composition of the transformations
\begin{equation}
  \label{NLDB}
  \hat F \to \hat F + \frac{ \delta V_F}{\delta F}=\hat F
  +\frac{\Delta F}{\grad F^2} ;\qquad (\hat F,t) \to (-\hat F, T-t)
\end{equation}
is thus a symmetry of the action.  Note the analogy with (\ref{DB1},\ref{DB2}).
The new variables read:
\begin{equation}
\label{NLDBTR}F_{\mbox{\tiny TR}}(x,t)=F(x,-t) \;\;\; ; \;\;\;
\hat F_{\mbox{\tiny TR}}(x,t)=-\hat F(x,-t)
  +\frac{\Delta F}{\grad F^2}(x,-t)
\end{equation}
Once again, rather than looking for excursions in the variables $F,\hat
F$, the problem is reduced to searching relaxations in the
variable $F_{\mbox{\tiny TR}},\hat F_{\mbox{\tiny TR}}$.

Because the system is driven out of equilibrium, it would be natural
to expect, as in section \ref{sec:DBR}, that this symmetry is
broken. {\em Remarkably, we shall show below that this symmetry is not
violated by the spatial boundary conditions, in spite of the system
being driven.} Transformation \eref{NLDB} indeed maps the
spatial boundary condition \eref{eqn:boundariesF} into
\begin{equation}
  \label{eqn:newBCF}
  F_{\mbox{\tiny TR}}(0)=\rho_0;\quad F_{\mbox{\tiny
      TR}}(1)=\rho_1;\qquad \hat F_{\mbox{\tiny
      TR}}(0)-\left.\frac{\Delta F_{\mbox{\tiny TR}}}{(\grad
    F_{\mbox{\tiny TR}})^2}\right|_{x=0}\!\!\!\!\!\!\!=\hat F_{\mbox{\tiny
      TR}}(1)-\left.\frac{\Delta F_{\mbox{\tiny TR}}}{(\grad
    F_{\mbox{\tiny TR}})^2}\right|_{x=1}\!\!\!\!\!\!\!=0
\end{equation}

Contrary to what happened in section \eref{sec:DBR}, the `zero noise
solution' $\dot F_{\mbox{\tiny TR}}= \Delta F_{\mbox{\tiny TR}}/2;\,
\hat F_{\mbox{\tiny TR}}=0$ this time satisfies
\eref{eqn:newBCF}. From \eref{eqn:boundariesF}, one indeed sees that
\begin{equation}
  \label{eqn:BCEQ1}
  \dot F|_{x=0,1} = \hat F|_{x=0,1} = 0
\end{equation}
Together with \eref{eqn:motionF}, this shows that any extremal
trajectory satisfies
\begin{equation}
  \label{eqn:BCEQ2}
\Delta F|_{x=0,1}=2[\dot F+\hat F
(\grad F)^2]|_{x=0,1}=0
\end{equation}
Furthermore, $\hat F_{\mbox{\tiny TR}}=0$ reads in the initial $F,\hat
F$ variables
\begin{equation}
  \label{eqn:instantonF}
  \hat F - \frac{\Delta F}{(\grad F)^2}=0
\end{equation}
which is indeed compatible with \eref{eqn:BCEQ1} and
\eref{eqn:BCEQ2}. Zero-noise diffusive relaxations in the variables
$F_{\mbox{\tiny TR}}, \hat F_{\mbox{\tiny TR}}$ thus correspond to the
excursion in the initial variables. The action of such a solution is
\begin{equation}
  S_F = [V_F]_0^T=\Big[\int \rmd x \log \grad F\Big]_0^T
\end{equation}
and the corresponding large deviation function is given, in the original
variables,  by
\begin{equation}
  \label{eqn:actioninstanton}
  S[\rho^*]= \int \rmd x \left[\rho \log
    \frac{\rho}{F}+(1-\rho)\log\frac{(1-\rho)}{1-F}+\log\grad F\right]_0^T
\end{equation}
$F$ is determined from $\rho$ by solving the equation $\hat
F_{\mbox{\tiny TR}}=0$ which reads
\begin{equation}
  \label{eqn:instantonrho}
  \rho=F + F(1-F) \frac{\Delta F}{\grad F^2}
\end{equation}
Injecting the temporal boundary conditions in
\eref{eqn:actioninstanton}, one gets
\begin{equation}
  \label{eqn:LDF}
  S[\rho^*]= \int \rmd x \left[\rho^* \log
    \frac{\rho^*}{F}+(1-\rho^*)\log\frac{(1-\rho^*)}{1-F}+\log\frac{\grad
      F}{\rho_1-\rho_0}\right]
\end{equation}

This is indeed the solution of the problem, initially found by
Derrida, Lebowitz and Speer~\cite{Derrida2001} and later recovered by
BDGJL \cite{Bertini2001}. Let us stress however that the
existence of a potential functional $V_F$ is a mystery, as nothing
guarantees that such a function exists out of equilibrium. Yet another
mystery is the fact that the symmetry related to $V_F$ is unbroken by
the reservoirs. We shall show below that these surprises are deeply
related to the fact that this model can be brought back to
equilibrium.

\pagebreak

\section{Non-local mapping to equilibrium}
\label{sec:BackToEq}

In the previous section we have shown that if one rotates the axes,
and expresses everything in terms of variables $F,\hat F$ associated
with the spin operators in the coherent state representation (or, in the
classical limit, with the stereographic projection of the spins), a
new, surprising detailed-balance symmetry becomes explicit. It is
unrelated to the original one and is not broken by the source terms at
the boundaries.  In this section we show that this sequence of
miracles can be condensed into only one: at the hydrodynamic level the
chain driven out of equilibrium can be mapped {onto a free equilibrium
chain with no sources at the boundaries}. The transformation that does
this is, however, non-local in space, and is the analogue of the one
we discussed in the introduction for non-interacting particles in a
potential.

Starting from the action $S_F$ \eref{eqn:actionFF}, one introduces the
{\em non-local} variables
\begin{equation}
  \label{eqn:primedef}
  \hat F'=  \nabla F;
  \qquad \grad F' = \hat F-\frac{\Delta F}{(\grad F)^2}
\end{equation}
where the second equation can also be written:
\begin{equation}
\hat F = \nabla\left[F'-\frac 1 {\hat F'}\right]
              = \nabla F' + \frac {\nabla \hat F'}{\hat {F'}^2}
\end{equation}
This change of variables takes the action into itself, apart from temporal
  boundary terms:
\begin{eqnarray}
  \label{eqn:Fprime}
  \!\!S\!=\!\!\int\!\!\text{d}x\!\left[ \rho \log\frac \rho F +
    (1-\rho) \log\frac
    {1-\rho}{1-F}+\log \hat F' -F' \hat F'\right]_0^T\nonumber\\
  + \int \text{d}x \text{d}t \left\{ \hat {F'} \dot F' +\frac
  1 2 \hat {F'}^2 (\nabla F')^2 +\frac 1 2 \nabla F' \nabla \hat
  F'\right\}
\end{eqnarray}
{Equation \eref{eqn:primedef} thus describes a non-local
  symmetry of the Hamiltonian}. This suggests that we continue the
succession of changes of variables done up to here
\begin{equation}
  (\rho,\hat \rho) \to (F, \hat F) \to (F',\hat F')
\end{equation}
by a further transformation $(F',\hat F') \to (\rho',\hat \rho')$,
where $\rho'$ and $\hat\rho'$ are related to $F',\hat F'$ in
the same way as are the
unprimed counterparts of \eref{eqn:rhotoF}
\begin{equation}
  \rho'= F' + F' (1-F') \hat F';\qquad \hat \rho' =
    \log\left(1+\frac{\hat F'}{1-F' \hat F'}\right)
\end{equation}
One thus ends up with
\begin{eqnarray}
  \label{eqn:mapaction}
\!\!\!\!\!\!\!\!\!    S\!=\!\!\!\int\!\!\text{d}x\!\left[\rho
 \log\frac \rho F + (1-\rho) \log\frac
    {1-\rho}{1-F}+\log \nabla F
    - \frac{\rho'-F'}{1-F'}-\rho' \log\frac
    {\rho'} {F'} - (1-\rho') \log\frac
    {1-\rho'}{1-F'}\right]_0^T\\
    +S'
\end{eqnarray}
where $S'=\int \text{d}t \text{d}x \left\{ \hat \rho' \dot \rho ' -
{\cal H}'\right\}$ and $ {\cal H}'$ is formally equivalent to the
initial Hamiltonian density $\HH$ but for the primed variables:
\begin{equation}
  S'=\int \text{d}t \text{d}x \left\{ \hat \rho' \dot \rho ' - 
  \frac 1 2 \sigma_{\rho'} \nabla \hat {\rho '}^2 + \frac 1 2 \nabla
  \rho ' \nabla \hat \rho'\right\}
\end{equation}
The overall change of variables, which reads
\begin{eqnarray}
  \label{eqn:nutshell}
  \nabla \left[\frac 1 {1-e^{\hat
  \rho'}}\right]&=(1-\rho)(e^{\hat\rho}-1)-\rho (e^{-\hat\rho}-1)\\
  \nabla \left[ \frac{\rho}{\rho + (1-\rho) e^{\hat \rho}}
  \right]&=(1-\rho')(e^{\hat\rho'}-1)-\rho' (e^{-\hat\rho'}-1)
\end{eqnarray}
thus maps the action of the hydrodynamic limit of the SSEP into
another SSEP. 

What we shall now prove is that this new process corresponds to an
isolated chain, and 
consequently possesses a detailed balance relation, induced by its
equilibrium entropy, which is not violated by the boundaries. Last, we
shall show that this detailed balance relation, mapped back to the
original $\rho, \hat \rho$ variables is the non-local symmetry
\eref{NLDB}.

\subsection{Boundary conditions - Currents}

From \eref{eqn:primedef} one sees that the spatial boundary conditions
\eref{eqn:boundariesF} read in the new variables
\begin{equation}
  \label{boundaryprime}
  \nabla\left[F'-\frac 1 
    {\hat F'}\right]_{x=0,L}=0;\qquad \int_0^L \hat F'=\rho_1-\rho_0
\end{equation}
{In the language of spin variables, the hydrodynamic Hamiltonian
\eref{eqn:hamiltonianspin} presents three conserved quantities
in the bulk - the components of the spins - which can be written}
\begin{equation}
Q_1=2\rho'- 1=2S'_z \quad; \quad   Q_2=\hat F'=2(S'_z+iS'_y) \quad; \quad
Q_3= \hat F'(1-2F')=2S'_x-1
\label{charges}
\end{equation}
Their continuity equations  read
\begin{equation}
  \dot Q_i= -\grad J_i
\end{equation}
where the currents $J_i$ are given by
\begin{eqnarray}
  &J_{Q_1}=- \nabla \rho' + 2 \sigma_{\rho'} \nabla \hat \rho'
  \;\;\;; \;\;\; J_{Q_2}= \frac 1 2 \nabla \hat F' + \hat F^{'2}
  \nabla F'\!\!\!\!\nonumber\\ &J_{Q_3}\!=\! (1-2F') \frac {\nabla \hat
  F'}2\!+\![\hat F'\!\!+\!{\hat {F'}}^{\!2}\!(1-2F')]\nabla F'\!\!\!\!
    \label{currents}
\end{eqnarray}

{Let us show that these currents vanish at the boundaries for all
  extremal trajectories. Such trajectories satisfy the equation of
  evolution \eref{eqn:motionF} and the boundary condition
  \eref{eqn:boundariesF}. This implies that $\Delta F$ vanishes at the
  boundary (l.h.s. o f \eref{eqn:motionF} together with $\hat
  F=0$). From the definition \eref{eqn:primedef} one then gets that
  $\grad \hat F'$ also vanishes, which implies, together with the
  l.h.s. of (\ref{boundaryprime}), that $\nabla F'$ also
  vanishes. Last, from the mapping $F',\hat F'\to \rho',\hat \rho'$ one
  sees that if both $\grad F'$ and $\grad \hat F'$ vanish, so do
  $\grad \rho'$ and $\grad \hat \rho'$. All in all, one gets that at the
  boundaries
\begin{equation}
  \grad F'=\grad \hat F'=\grad \rho'=\grad\hat\rho'=0
\end{equation}
We then see from \eref{currents} that the three currents $J_{Q_i}$
vanish at the boundary: {\em the transformed model in the primed
  variables is an isolated chain}. This condition alone, supplemented
with the r.h.s. of (\ref{boundaryprime}), encompasses all the original
boundary conditions.}

\subsection{Profiles and trajectories}

When the dust sets, we see that all that has been used is the fact that
the original variables $(\rho,\hat \rho)$ or, more physically $(\rho,J)$,
have been mapped by a non-local transformation into new densities and currents
\begin{equation}
(\rho(x,t),J(x,t)) \;\;\; \rightarrow \;\;\; (\rho'(x,t),J'(x,t))
\label{mapp}
\end{equation}
The original detailed balance symmetry (\ref{ddbb}) that maps a
trajectory into its time-reversed:
\begin{equation}
(\rho(x,t),J(x,t)) \;\;\; \rightarrow \;\;\; (\rho(x,-t),-J(x,-t))
\end{equation}
 is broken by the source terms driving the system out of equilibrium.
Miraculously, the transformed chain is isolated, so that time-reversed 
trajectories are related through:
\begin{equation}
(\rho'(x,t),J'(x,t)) \;\;\; \rightarrow \;\;\; (\rho'(x,-t),-J'(x,-t))
\end{equation}
Coming back to the original variables via (\ref{mapp}) mixes the density $\rho'$
(symmetric in time) with the current $J'$ (antisymmetric in time), thus making
the pair of transformed trajectories expressed in  
$\rho$ and $J$ neither  symmetric nor antisymmetric.

Let us now describe `uphill' and `downhill' trajectories. 
The stationary profile $\bar \rho$
maps to a flat profile $\bar \rho' =\text{C}^{st}$. The precise value 
of $\text{C}^{st}$ is arbitrary, due to dilation-invariance of
the model, contrary to $\hat F'$ which is constrained by the r.h.s of
(\ref{boundaryprime}). 

\begin{itemize}

\item {\bf Relaxations}: diffusive trajectories
of the initial model satisfy
\begin{equation}
  \dot \rho=\frac 1 2 \Delta\rho,\qquad\,\hat \rho=0
\end{equation}
From \eref{eqn:nutshell}, one sees that $\hat \rho=0$ implies $\grad
\hat \rho'=0$. As the primed variables also satisfy the equations of
motion \eref{classicalequations1}, the resulting trajectories evolve
with
\begin{equation}
  \dot \rho'=\frac 1 2 \Delta \rho'
\end{equation}
{\em Relaxations thus map into relaxations}. 

\item {\bf Excursions}: The instanton equations \eref{eqn:instantonF}
imply $\nabla F'=0$ (cfr (\ref{eqn:primedef})). Using the relation
of $F, \hat F$ \eref{eqn:rhotoF}, as applied to the primed variables,
 one gets
\begin{equation}
  \nabla \hat \rho'-\frac{\nabla \rho'}{\sigma_{\rho'}}=0
\end{equation}
Injected back in the equations of motion \eref{classicalequations1},
this shows that densities evolve with 
\begin{equation}
  \dot \rho'=-\frac 1 2 \Delta\rho'
\end{equation}
{\em Excursions of the initial model, once mapped back to equilibrium
  through \eref{eqn:nutshell}, are given by time-reversal of the
  isolated chain's relaxations.} The action $S'[\hat \rho',\rho']$ of
such an uphill trajectory is $\int_0^L \text{d} x [\rho' \log\rho'
  +(1-\rho')\log(1-\rho')]_0^T$.  As $F'$ is constant along the
instanton and $\int_0^L \rho'$ is a constant of motion, the overall
action (\ref{eqn:mapaction}) reduces to the large deviation function
\eref{eqn:LDF}, as it should.

\end{itemize}

\subsection{Detailed balance symmetry}

Let us now show in terms of spin variables how the detailed balance
relation of the isolated chain accounts for the miraculous
transformation \eref{NLDB} which allowed BDGJL to compute the
large deviation function. In the spin variables, the original
detailed balance symmetry \eref{DB1} and \eref{DB2} amounts to a
reflexion of all the spins with respect to the $x-z$ plane:
\begin{equation}
  \label{eqn:DBSpins}
  (S_x,S_y,S_z)\to (S_x,-S_y,S_z);\qquad T\to T-t
\end{equation}

Because the bulk 
Hamiltonian  is also invariant with respect to
 any simultaneous  rotation of all the
spins, any composition of \eref{eqn:DBSpins} with a rotation gives
another `detailed-balance like' symmetry. These symmetries are all
broken by the boundary conditions of the original model.
Once mapped to the isolated chain, the boundary conditions
\eref{boundaryprime} reduce to fixing the value of an integral of
motion:
\begin{equation}
  \label{conservation}
  2\int_0^L (S_z'+i S_y')=\int_0^L \hat F'=\rho_1-\rho_0
\end{equation}
Among all the `detailed-balance like' symmetries of the isolated
chain, only one preserves \eref{conservation}. We are thus left
with the transformation
\begin{equation}
\label{eqn:dbisolspin}
  (S'_x,S'_y,S'_z)\to(-S'_x,S'_y,S'_z);\qquad t\to T-t
\end{equation}
From the expression of $Q_3$ in \eref{charges}, one sees that
${S_x^{\mbox{\tiny TR}}}'=-S_x'$ can be written as
\begin{equation}
  1+\hat F_{\mbox{\tiny TR}}'(1-2 F'_{\mbox{\tiny TR}})=-1-\hat F'(1-2 F')
\end{equation}
Using $\hat F'=\grad F$ and looking for $F_{TR}=F$, this reads
\begin{equation}
  F'+F'_{\mbox{\tiny TR}}=1+\frac 1 {\grad F}
\end{equation}
Differentiating once and using $\grad F'=\hat F-\frac{\Delta F}{(\grad
F)^2}$, one gets
\begin{equation}
  \hat F_{\mbox{\tiny TR}}+\hat F=\frac{\Delta F}{(\grad F)^2}
\end{equation}
which is nothing but the non-local mapping (\ref{NLDBTR}) between
`downhill' diffusive solutions and the instantons. One thus sees that
the miraculous solution of the initial model is simply induced by the
detailed balance-like relation \eref{eqn:dbisolspin} of the isolated
chain, which does not violate the boundary conditions
\eref{boundaryprime}.

\pagebreak

\section{KMP}
\label{sec:KMP}

The Kipnis-Marchioro-Presutti model (KMP) was introduced
in~\cite{KMP1982} as a one dimensional model of energy transport
satisfying Fourier Law. Bertini, Gabrielli and Lebowitz
 recently computed the large
deviation function of the energy profile using an approach similar to
the one used previously for the SSEP~\cite{Bertini2005}. We shall
show below that once again a mapping back to equilibrium explains this
success.

The functional expression for the fluctuating hydrodynamics of KMP is
very similar to that of SSEP:
\begin{equation}
  \label{eqn:action}
  \int {\cal D}[\hat \rho,\rho] e^{-N S[\hat \rho, \rho]}=\int
  {\cal D}[\hat \rho,\rho]e^{- N \int\text{d}t \text{d}x \{ \hat
    \rho \dot \rho - {\cal {H}}\}}
\end{equation}
where we have introduced the Hamiltonian density:
\begin{equation}
\label{eqn:hamdens}
{\cal{H}}\equiv \frac{1}{2} \left[ \rho^2 \nabla \hat \rho^2
- \grad \hat \rho \grad \rho\right]
\end{equation}
To compute the large deviation function ${\cal F}(\rho^*)$, one has to
solve the corresponding Hamilton equations
\begin{equation}
  \label{eqnmotionKMP}
  \dot \rho = \frac 1 2 \Delta \rho-\grad[\rho^2 \grad \hat\rho] ; \qquad\dot {\hat \rho}=-\frac 1 2 \Delta \hat \rho -\rho (\grad \hat\rho)^2
\end{equation}
with the boundary conditions
\begin{eqnarray}
\label{eqn:KMPBC}
    \rho(x,0)&=\bar \rho(x)=(1-x)\rho_0+x 
    \rho_1;\quad \rho(x,T)=\rho^*(x)\\
    \rho(0,t)&=\rho_0;\quad
    \rho(1,t)=\rho_1;\quad  \hat \rho(0,t)= \hat \rho(1,t)=0
\end{eqnarray}
Once again, the last equality simply says that no fluctuations are
allowed at the contact with the
reservoir.

\subsection{Connection with the SU(1,1) spin chain}
KMP is related to SU(1,1) spin chains in a rather subtle
way~\cite{Giardina2006,Giardina2008}. Starting from the SU(1,1)
coherent states for spin $k$
\begin{equation}
  |z\rangle = \frac{1}{(1-z \bar z)^k}e^{z K^+} |0\rangle
\end{equation}
one gets the following expression for the pseudo-spin operators
\begin{equation}
  \label{eqn:spinopsu11}
  \langle z | K^+ | z\rangle= 2k\frac { \bar z}{1-z \bar z};
  \qquad 
  \langle z | K^- | z\rangle =2k \frac{z}{1-z \bar z};
  \qquad 
  \langle z | K^z | z\rangle = k\frac{1+z \bar z}{1-z\bar z}
\end{equation}
of the SU(1,1) group. By analogy with equation \eref{eqn:doipelitiSU2bis}
for the SU(2) case, the Doi-Peliti variables are defined through
\begin{equation}
  \label{eqn:doipelitiSU11}
   \phi^*=\bar z;\qquad \phi=\frac{z}{1-z\bar z}
\end{equation}
The difference with the SU(2) case stems from the fact that the energy
density variables $\rho, \hat \rho$ have to be identified directly
with the Doi-Peliti variables:
\begin{equation}
  \label{rhoetlreste}
  \rho =\phi=\frac{z}{1-z\bar z};\qquad \hat \rho=\phi^*=\bar z
\end{equation}
In particular, from \eref{eqn:spinopsu11} and \eref{rhoetlreste}, one
 sees that $\rho$ corresponds to $K^-$ and not to the $z$ component of
 the spins as was the case for the SSEP and the SU(2)
 representation~\cite{foot3}. Defining the classical spin
\begin{equation}
  K^{x,y,x} = \frac 1 {2k} \langle z| K^{x,y,z} | z \rangle
\end{equation}
one then checks that the Hamiltonian \eref{eqn:hamdens} corresponds to
 the continuous pseudo-spin chain
\begin{equation}
\label{eqn:hamSu11}
\HH= -\frac 1 2 \left[ (\grad K_x)^2+(\grad K_y)^2-(\grad K_z)^2\right]= -\frac 1 2 \left[ \grad K^+ \cdot \grad K^--(\grad K_z)^2\right]
\end{equation}

\subsection{Remarkable change of variables}
As we did in section \ref{sec:FetHatF} for the SSEP, we shall use the
symmetry of the evolution operator under simultaneous SU(1,1) `rotation'
of all `spins'  to find a basis where a
non-local mapping to equilibrium is easily revealed.
Making a reflexion with respect to the $x-z$ plane maps $K^y$ in
$-K^y$ and lets \eref{eqn:hamSu11} invariant. As expected, it is thus a
symmetry of the action. In the $(z,\bar z)$ coordinates it reads
\begin{equation}
  z'=\bar z;\qquad \bar z'=z
\end{equation}
By analogy with \eref{eqn:doipelitiSU11}, one defines
\begin{equation}
  \phi'=\frac{z'}{1-z'\bar z'};\qquad  {\phi^*}'=\bar z'
\end{equation}
The transformation then reads in density variables
\begin{equation}
  \label{phirho}
  \rho= {\phi^*}' (1+\phi  {\phi^*}');\quad \hat \rho =
  \frac{\phi'}{1+\phi' {\phi^*}'};\quad \leftrightarrow\quad
  \phi'=\hat \rho (1+\rho \hat \rho);\quad {\phi^*}' =
  \frac{\rho}{1+\rho \hat \rho}
\end{equation}
As for the SSEP, we could now work with these new 'Doi-Peliti' like
variables, but to make contact with the solution of BDGJL we
use slightly different notations
\begin{equation}
  F={\phi^*}';\qquad \hat F=\phi'
\end{equation}
so that the overall mapping of the action reads
\begin{eqnarray}
  \label{eqn:KMPactionF}
  S[\rho,\hat \rho]&=  \int \rmd x \left[ \frac \rho  F-\log \frac \rho F
    \right]_0^T \\
  &+
  \int \rmd x \rmd t \left\{ \hat F \dot F - \frac 1 2 \hat F ^2 (\grad F)^2
  + \frac 1 2 \grad \hat F \grad F \right\}
\end{eqnarray}
The fields $F,\hat F$ are related to energy densities through
\begin{equation}
\label{eqn:KMPF}
  F=\frac{\rho}{1+\rho\hat \rho};\qquad \hat F=\hat\rho(1+\rho\hat\rho)
\end{equation}
so that the spatial boundary conditions \eref{eqn:KMPBC} are given by
\begin{equation}
  \label{eqn:FBC}
  \hat F(0)=\hat F(1)=0;\qquad F(0)=\rho_0;\qquad F(1)=\rho_1
\end{equation}

\subsection{Non-local mapping back to equilibrium}
Let us introduce the non-local variables
\begin{equation}
  \label{dualvarKMP}
  \hat F' = \grad F;\qquad \hat F = \grad \left[F'+\frac 1 {\hat F'}\right]
\end{equation}
which maps the action into
\begin{eqnarray}
  \label{dualF'}
  S[\rho,\hat \rho]&= \int \rmd x \left[ \frac \rho F-\log \frac \rho F
    -\log \hat F' - F' \hat F'\right]_i^f \\
  &+ \int \rmd x \rmd t \left\{ \hat F' \dot F' - \frac 1 2 \hat{ F'}^2 (\grad
  F')^2 + \frac 1 2 \grad \hat F' \grad F' \right\}
\end{eqnarray}
Comparison of \eref{dualF'} and \eref{eqn:KMPactionF} reveals that
\eref{dualvarKMP} is a non-local symmetry of the action. The boundary
conditions \eref{eqn:FBC} now reads
\begin{equation}
\label{BCprime}
\int_0^L \hat F' = \rho_L - \rho_0 \qquad \grad F' -\frac {\grad \hat
  F'}{\hat F'^2}=0
\end{equation}
As the classical trajectories in the original $F$, $\hat F$ variables
satisfy
\begin{eqnarray}
  \dot F &= \frac 1 2 \Delta F + \hat F \grad F^2\\
  \dot {\hat F} &= -\frac 1 2 \Delta \hat F + \grad \left[ \hat F^2
    \grad F\right]
\end{eqnarray}
we see that on the boundaries, $\dot F=0$ and $\hat F=0$ implies $\Delta
F = \grad \hat F'=0$. Together with the r.h.s. of (\ref{BCprime}), this
implies
\begin{equation}
  \grad F' = \grad \hat F' =0
\end{equation}
and all the currents vanish on the
boundaries. \eref{dualvarKMP} thus maps the chain into an isolated
one. One can continue the mapping $\rho,\hat \rho\to F,\hat F\to
F',\hat F'$ to introduce new energy densities $\rho',\hat \rho'$:
\begin{equation}
  \rho'=F' (1+F' \hat F');\qquad \hat \rho'= \frac{\hat F'}{1+F' \hat F'}
\end{equation}
One then gets an overall action
\begin{eqnarray}
  S[\rho,\hat \rho]&= \int \rmd x \left[ \frac \rho F-\log \frac \rho F
    -\log \hat F' - F' \hat F'-\frac {\rho'}{ F'} + \log \frac{\rho'}{F'}\right]_i^f \\
  &+ \int \rmd x \rmd t \left\{ \hat \rho ' \dot \rho ' - \frac 1 2 {\rho
    '}^2 (\grad \hat \rho')^2 + \frac 1 2 \grad \hat \rho ' \grad
  \rho '\right\}
\end{eqnarray}
which shows that the hydrodynamic limit of the KMP model
driven out-of-equilibrium can be mapped back to equilibrium through a
non-local change of variables. This chain has a detailed balance
symmetry, and as before its instantons are time-reversal of
relaxations and are thus given by
\begin{equation}
\dot \rho'=-\frac 1 2 \Delta \rho'
\end{equation}
From the equations of motion \eref{eqnmotionKMP}, one sees that this
corresponds to
\begin{equation}
  \rho^2 \grad \hat \rho=\grad \rho
\end{equation}
In the $F',\hat F'$ variables, this reads
\begin{equation}
  \grad F'=0
\end{equation}
Mapped back to the initial variables, one gets the instanton equation
\begin{equation}
  \hat F = -\frac{\Delta F}{(\grad F)^2}
\end{equation}
or, for the density variables,
\begin{equation}
  \rho = F - F^2\frac{\Delta F}{\grad F^2}
\end{equation}
This is the counterpart for KMP of the instanton equation
\eref{eqn:instantonrho} for the SSEP, and corresponds to the equation
found by BDGJL. The action of this trajectory is
\begin{equation}
  S[\rho,\hat \rho]= \int \rmd x \left[ \frac \rho F-\log \frac \rho F
    -\log \hat \grad F\right]_0^T
\end{equation}
which is precisely the large deviation function obtained
in~\cite{Bertini2005}.  One sees that once again a non-local mapping
back to equilibrium enables one to find the instanton equation and
thus compute the large deviation function. This result can be extended
to the whole class of systems defined by a Hamiltonian density
$H(\rho,\hat\rho) = \sigma  (\grad \hat \rho)^2/2 -\grad \rho \grad
\hat \rho/2$ where $\sigma$ is a second order polynomial in $\rho$.

\pagebreak
\section{Non-interacting particles 
in an arbitrary potential driven out-of-equilibrium}
\label{sec:duality_FP}

Apart from the simple example addressed in the introduction, the
mappings to equilibrium we have presented so far only apply at the
level of large deviations. We shall show below that such a mapping can
also be constructed for the full probability distribution, without
coarse-graining, of a model of {\em non-interacting} particles driven
out-of-equilibrium.  We shall do the mapping in two ways: at the level
of probabilities in subsection \ref{ooop}, and at the level of
evolution operators in subsection \ref{sec:TFEO}.
 
\begin{figure}[h]
  \begin{center}
    \includegraphics[width=.7\columnwidth]{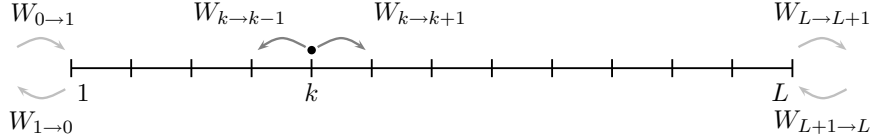}
    \caption{Open chain in contact with reservoirs: generic rates for
      individual particles.}
    \label{fig:dyn_W_gen}
  \end{center}
\end{figure}

We consider an open chain of $L$ sites (index $1\leq k\leq L$) in
contact with two reservoirs. Each particle at site $k$ can jump to a
neighboring site $k\pm 1$ with rates $W_{k\to k\pm 1}$.
The system is coupled to reservoirs at the two boundaries (sites $1$ and $L$) through
transition rates $W_{0\to 1},W_{1\to 0},W_{L\to L+1},W_{L+1\to L}$
(See figure~\ref{fig:dyn_W_gen}).

%
%
A simple case for which the steady state is known is that of
equilibrium systems, where probability currents vanish. For
non-interacting particles, the corresponding detailed balance relation
is equivalent to balancing the stationary mass fluxes over each
bond. Introducing the average occupancies $P_k^\eq$, it reads:
\begin{equation}
  \label{eqn:DBbulkmicro}
  W_{0\to 1}= P_1^\eq W_{1\to 0}\qquad P^\eq_k \, W_{k\to k+1} =
  P^\eq_{k+1} \, W_{k+1\to k} \qquad W_{L\to L+1} P_L^\eq = W_{L+1\to L}
\end{equation}
The first two equations imply that $P^\eq_k$ takes the form:
\begin{equation}
  P^\eq_k= \prod_{\ell=0}^{k-1} 
  \frac{W_{\ell\to\ell+1}}{W_{\ell+1\to\ell}}
  \qquad\textnormal{for}\qquad 1\leq k\leq L
  \label{eq:def_Peq}
\end{equation}
Combining the last equation of \eref{eqn:DBbulkmicro} and
\eref{eq:def_Peq}, we see that this solution is consistent as long as
\begin{equation}
  \label{eqn:DBglobcond}
  \frac{W_{0\to 1} W_{1\to 2} \dots W_{L\to L+1}}{W_{L+1\to L} \dots
    W_{1 \to 0}}=1
\end{equation}

This condition can be violated in many physical situations as for
instance when a current is forced by the reservoirs. In such cases,
the steady state occupancies are not the $P^\eq_k$'s and one has to
resort to other methods for their determination,
as has been done for instance by Derrida in~\cite{Derrida1983}
for a particle hopping in a periodic potential. To our knowledge, the
case of open systems with non-vanishing steady current has not been
considered in the literature and could be interesting in the context
of non-equilibrium disordered media~\cite{Bouchaud1990}.

In the following section, we show how to map our open system back to
equilibrium, allowing us to get an explicit expression for the steady
state distribution.

\subsection{Transformation of probabilities}
\label{ooop}

Let us consider a probability distribution obtained as a product of
Poisson distribution in each site
\begin{equation}
  \label{eqn:factorized}
  P(n_1,\dots,n_L)=\prod_{k=1}^L \frac{P_k^{n_k}}{n_k!} e^{-P_k}
\end{equation}
Its form is preserved by the time evolution and the $P_k$'s evolve with the conservation equation
\begin{equation} 
  \label{eqn:partcurrent}
  \partial_t P_k = -(J_{k+1}-J_k)
\end{equation}
where one has  introduced the currents
\begin{eqnarray}
  \label{eq:prob_conserv_current}
   J_k=-P_k W_{k\to k-1} + P_{k-1} W_{k-1\to k} \qquad (\text{for } 2\leq k\leq L)\\
   J_1=-P_1 W_{1\to 0} + W_{0\to 1} \qquad
   J_{L+1}=-W_{L+1\to L}+P_L W_{L\to L+1}
\end{eqnarray}

In general, the steady state is not given by cancelling all currents
$J_k$, as this takes us back to the detailed balance
conditions~\eref{eqn:DBbulkmicro}. In the spirit of previous sections,
our aim here is to map our model to an isolated system, for which
currents vanish at the boundaries.  The construction of the
equilibrium model follows closely the one of the Fokker-Planck
equation presented in the introduction. We introduce primed
occupancies
\begin{eqnarray}
  \label{eqn:defPprime}
  P_k'=(P_{k+1}^\eq)^{-1}P_{k+1}-(P_k^\eq)^{-1}P_k; \qquad (\text{for } 1\leq k\leq L-1)\\
  P'_0 = (P_1^\eq)^{-1} P_1 - \big(P^\eq_0\big)^{-1};\qquad P'_L = \big(P^\eq_{L+1}\big)^{-1}-(P_L^\eq)^{-1}P_L\nonumber
\end{eqnarray}
where one has extended the definition \eref{eq:def_Peq} of $P_k^\eq$
to $P_{L+1}^\eq$ and set $P_0^\eq=1$. Defining the new rates
\begin{equation}
  W'_{k\to k+1}=W_{k+1\to k};\qquad W'_{k+1\to k}=W_{k+1\to k+2}
  \label{eq:primed_rates_W}
\end{equation}
we now have the evolution for a chain of $L+1$ sites ($0\leq k \leq L$)
\begin{equation}
  \label{eq:bulk_primed_mastereq}
  \partial_t P'_k =- (J_{k+1}'-J_k')
\end{equation}
where the currents are defined by
\begin{equation}
  J_0'=J_{L+1}'=0; \quad\text{and}\quad J'_k = -P'_kW'_{k\to
  k-1}+P'_{k-1}W'_{k-1\to k};\qquad (\text{for } 1\leq k\leq L)
\end{equation}
Strikingly, the currents $J_0'$ and $J_{L+1}'$ vanish and the primed
chain is thus isolated.

Note that although we mapped our initial open chain of $L$ sites into
an isolated chain of $L+1$ sites, there is no contradiction when
counting degrees of freedom. Indeed, from the
definition~(\ref{eqn:defPprime}), we see that the primed occupancies satisfy
\begin{equation}
  \sum_{k=0}^LP'_k  \ = \
  \big(P^\eq_{L+1}\big)^{-1} - \big(P^\eq_0\big)^{-1}
  \label{eq:normPprime}
\end{equation}
which means that there are only $L$ independent occupancies in the
isolated chain.

The (equilibrium) steady state of the primed chain is found by
imposing $J'_k=0$ for all $k$'s:
\begin{equation}
  P'_k \propto P'_\eq(k) \ = \
  \prod_{\ell=0}^{k-1} 
  \frac{W'_{\ell\to\ell+1}}{W'_{\ell+1\to\ell}}
  \ = \
  \frac{W_{0\to 1}}{W_{k+1\to k}P^\eq_{k+1}}
  \ = \
  \frac{W_{0\to 1}}{W_{k\to k+1}P^\eq_{k}}
\label{eq:primed_Peq}
\end{equation}
Mapping back to the original variables, one obtains the expression
of the steady state occupancies $P_k^{\text{st}}$
\begin{equation}
  \label{eq:res_Pk}
  P_k^{\text{st}}=\frac{1}{Z_L} \left[(P_0^\eq)^{-1} \sum_{\ell=k}^L \frac {P_k^\eq}{W_{\ell \to \ell+1} P_\ell^\eq}+(P_{L+1}^\eq)^{-1} \sum_{\ell=0}^{k-1} \frac {P_k^\eq}{W_{\ell \to \ell+1} P_\ell^\eq}\right]
\end{equation}
where $Z_L$ is a normalization constant given by
\begin{equation}
  Z_L=\sum_{\ell=0}^{L} \frac {1}{W_{\ell \to \ell+1} P_\ell^\eq}
\end{equation}
The result \eref{eq:res_Pk} is reminiscent of that obtained by Derrida
in~\cite{Derrida1983}, the differences highlighting the role of the
reservoirs.

To highlight the connection with the case treated in introduction, let
us consider explicitly particles diffusing in a discrete potential,
where the rates $W_{k\to k\pm 1}$ are given by
\begin{equation} \label{eq:unprimedratesV}
  W_{k\to k\pm 1} = \beta^{-1}\,e^{-\frac 12\beta (V_{k\pm 1}-V_k)}  
\end{equation}
In the bulk, they obey detailed balance with respect to the Boltzmann
weight $P_k^{\text{eq}}\propto e^{-\beta V_k}$, but the system is
driven out-of-equilibrium as soon as $V_{L+1}\neq V_{0}$ (see
relation~\eref{eqn:DBglobcond}). The averaged occupancies $P_k$
evolve with
\begin{equation} \label{eq:eqM_exp-rates}
  \beta \partial_t P_k = P_{k+1} e^{-\frac 12\beta (V_{k}-V_{k+1})} +
  P_{k-1} e^{-\frac 12\beta (V_{k}-V_{k-1})} -P_k \,\big[ e^{-\frac
      12\beta (V_{k+ 1}-V_k)} + e^{-\frac 12\beta (V_{k- 1}-V_k)}
    \big]
\end{equation}
Taking the continuum limit ($k=x L$ with $0\leq x\leq 1$, $L\gg 1$)
and rescaling time with $L^2$ (diffusive scaling), one recovers the
Fokker-Planck equation~\eref{eq:FPcontinuum} with $\beta^{-1}=T$,
upon gradient expansion of~\eref{eq:eqM_exp-rates}.
In particular, taking the continuum limit of the microscopic steady
state~\eref{eq:res_Pk}, one recovers as expected the
result~\eref{eq:Pstat_FP} for the non-equilibrium Fokker-Planck steady
state.
Last,  with the potential $V_k$ defined
in~\eref{eq:unprimedratesV}, the primed equilibrium law reads
$P'_\eq(k)\propto \exp\{\frac \beta 2(V_k+V_{k+1})\}$, which is analogous to the
sign change of $V$ for Fokker-Planck (see
equation~\eref{eq:evolPprimed_FP}), but also induces a smooth
averaging of the potential over two neighboring sites.

\subsection{Transformation for evolution operators}
\label{sec:TFEO}

As shown above, the average occupancies $P_k$ fully determine the
steady state of non-interacting particles and can be computed through
a mapping to equilibrium. Such approach does not apply for exclusion
processes, as exclusion correlates particles. To cast the mapping to
equilibrium in a form similar in both cases, we shall now work
directly with the evolution operator of non-interacting particles.
Beyond the quest for a formalism ultimately applicable to microscopic
models of interacting particles, this point of view proves to be much
stronger than the one developed in section \ref{ooop}, even for
non-interacting particles. Indeed, it applies when the initial
distribution is not factorized, i.e. not in the form
\eref{eqn:factorized}, and gives access not only to the steady state,
but also to the whole dynamics.

For the system of non-interacting particles considered in this
section, the probability $P(\boldsymbol n,t)$ evolves with (we refer
to section \ref{sec:mastereqandspin} for the notation):
\begin{eqnarray}
  \partial_t P(\boldsymbol n,t) =  \sum_{k=1}^{L-1} \Big\{ \
  n_{k+1}^+ W_{k+1\to k} P(n_k^-,n_{k+1}^+) + n_{k}^+ W_{k\to k+1}
  P(n_k^+,n_{k+1}^-) - \big[ n_k W_{k\to
      k+1}\nonumber\\\qquad\qquad + n_{k+1} W_{k+1\to k} \big]\: P(\boldsymbol n)\ \Big\}
  + W_{1\to 0} \Big[ n_1^+ P(n_1^+) - n_1
  P(\boldsymbol n) \Big] + W_{0\to 1} \big[P(n_1^-) - P(\boldsymbol
    n)\big] \nonumber \\ \qquad\qquad + W_{L\to L+1} \Big[ n_L^+ P(n_L^+) - n_L
  P(\boldsymbol n) \Big] + W_{L+1\to L}\big[P(n_L^-) - P(\boldsymbol
  n)\big]
\label{eq:evol_P_occupnb}
\end{eqnarray}
Let us introduce the usual Doi-Peliti~\cite{Doi1976,Peliti1985}
creation and annihilation operators $a_k^\dagger, a_k$, defined by
\begin{equation}
  a_k^\dagger |n_k\rangle
  =|n_k+1\rangle;\qquad a_k
  |n_k\rangle =n_k
  |n_k-1\rangle
\end{equation}
The master equation can then be written as $\partial_t |\psi\rangle =
- \mathbb H |\psi\rangle$, where $|\psi\rangle=\sum_{\bol n} P(\bol n)
|\bol n\rangle$, and the evolution operator $\mathbb H$ reads
\begin{eqnarray}
  \mathbb H &= 
   \sum_{k=1}^{L-1}  (a^\dag_{k+1} - a_k^\dag) \big\{ a_{k+1} W_{k+1\to k} - a_k W_{k\to k+1} \big\}
\nonumber \\ &\qquad -
  (a_1^\dag-1)\big(  W_{0\to 1}  - W_{1\to 0} a_1 \big) \ - \ 
  (a_L^\dag-1)\big(  \, W_{L+1\to L} -  W_{L\to L+1}a_L   \big)
\label{eq:opWboundGeneral}
\end{eqnarray}
Because of the boundary terms, this operator does not correspond to an
equilibrium dynamics. We know however that its ground state can be
mapped to that of an equilibrium operator (see section
\eref{ooop}) and it is thus quite natural to investigate the question as
to whether this mapping extends to the whole operator.

\smallskip For the exclusion process, the equilibrium model was
constructed at macroscopic coarse-grained level through the use of
canonical transformations, which mapped the action onto that of an
equilibrium, isolated system. At the level of operators, canonical
transformations correspond to similarity transformations $\mathbb H'=
Q^{-1} \mathbb H Q$ (see \ref{sec:symmetryFromMicrotoMacro}). We will
now show that using such transformations, one can map $\mathbb H$ to
an equilibrium operator. As these transformations do not modify the
spectrum, the determination of the eigenstates reduces to the
determination of the spectrum of an equilibrium operator.

We first translate the $a_k^\dag$'s by $1$, using the transformation
obtained from $Q_0={e}^{-\sum_{k=1}^L a_k }$ (one has $Q_0^{-1} a_k^\dag Q_0 =
a_k^\dag+1$). Rearranging the sum, we get for $ \mathbb H_0=
Q_0^{-1}\mathbb H Q_0$
\begin{eqnarray}
 \mathbb H_0   &= - 
  \sum_{k=2}^{L-1}  a^\dag_k 
  \Big\{ 
    \big( W_{k+1\to k} a_{k+1}  - W_{k\to k+1} a_k \big) \:-\:
    \big( W_{k\to k-1} a_{k}  - W_{k-1\to k} a_{k-1} \big)
  \Big\}
\nonumber \\ &\qquad -
  a^\dag_1\Big\{
     \big( W_{2\to 1} a_{2}  - W_{1\to 2} a_1 \big) \:-\:
     \big(W_{1\to 0} a_1  - W_{0\to 1}  \big) \Big\}
\nonumber \\ &\qquad -
  a^\dag_L\Big\{
     \big( W_{L+1\to L}  -  W_{L\to L+1}a_L   \big) \:-\:
     \big( W_{L\to L-1} a_{L}  - W_{L-1\to L} a_{L-1} \big)
     \Big\}
\end{eqnarray}
Then, we use the similarity transformation induced by
$Q_1=\prod_{k=1}^L \big(P_k^{\text{eq}}\big)^{a^\dag_k a_k}$, which yields
\begin{equation}
   \cases{
            Q_1^{-1} a_k Q_1 =  P_k ^{\textnormal{eq}} a_k \\
            Q_1^{-1} a_k^\dag Q_1 = \big(P_k^{\textnormal{eq}}\big)^{-1} a_k^\dag
            }
\end{equation}
and thus for  $ \mathbb H_1= Q_1^{-1}\mathbb H_0 Q_1$ :
\begin{eqnarray}
 \mathbb H_1   &=  
-  \sum_{k=2}^{L-1}  a^\dag_k 
  \Big\{ 
    W_{k\to k+1} \: \big( a_{k+1} - a_k     \big) \:-\:
    W_{k\to k-1} \: \big( a_{k}   -  a_{k-1}\big)
  \Big\}
\nonumber \\ &\qquad -
  a^\dag_1\Big\{
     W_{1\to 2} \big( a_2  - a_1 \big) \:-\:
     W_{1\to 0} \big( a_1  -   (P^\eq_0)^{-1} \big) \Big\}
\nonumber \\ &\qquad -
  a^\dag_L\Big\{
     W_{L\to L+1} \big(   (P^\eq_{L+1})^{-1} -  a_L   \big) \:-\:
     W_{L-1\to L} \big( a_{L}  -a_{L-1} \big)
     \Big\}
\label{eq:expr_opW1}
\end{eqnarray}
Noticing that the boundary terms (second and third lines
of~\eref{eq:expr_opW1}) look like the bulk term (first line), with
$(P^\eq_0)^{-1}$ playing the role of an operator $a_0$, and
$(P^\eq_{L+1})^{-1}$ the role of an operator $a_{L+1}$, we add two
sites $k=0$ and $k=L+1$, with their corresponding creation and
annihilation operators $a^{(\dag)}_{0}$, $a^{(\dag)}_{L+1}$. We also
define two vectors $|L\rangle$ and $|R\rangle$, satisfying $
a_0|L\rangle = (P^\eq_{0})^{-1}\: |L\rangle \:,\ a_{L+1}|R\rangle =
(P^\eq_{L+1})^{-1}\: |R\rangle $~\cite{foot6}. Defining, on the
extended space (of $L+2$ sites)
\begin{eqnarray}
  \label{eqn:definitionH2}
  {\mathbb H}_2   &=  
  -\sum_{k=1}^L  a^\dag_k 
  \Big\{ 
    W_{k\to k+1} \: \big( a_{k+1} - a_k     \big) \:-\:
    W_{k\to k-1} \: \big( a_{k}   -  a_{k-1}\big)
  \Big\} 
\end{eqnarray}
we observe from~\eref{eq:expr_opW1} that
\begin{equation}
 {\mathbb H}_2\: \Big(
  \: |L\rangle \otimes |n_1,\ldots,n_L\rangle \otimes |R\rangle \:\Big)
  \ = \
   |L\rangle \otimes 
   \Big(\:{\mathbb H}_1 \:  |n_1,\ldots,n_L\rangle  \:\Big) 
   \otimes |R\rangle 
   \label{eq:W1W1tilde_vects}
\end{equation}
In other words, the action of the extended operator $ {\mathbb H}_2$
  on states of the form $|L\rangle \otimes |n_1,\ldots,n_L\rangle
  \otimes |R\rangle$ reduces to that of ${\mathbb H}_1$ on the
  physical state $|n_1,\ldots,n_L\rangle$.

A last transformation brings $\mathbb H_2$ into an equilibrium form; it
is generated by the non-local operator
\begin{equation}
  Q_2 = \exp\sum_{0\leq p< q \leq L+1}  \frac{a^\dag_p a_q}{q-p}
\end{equation}
which transforms the $(L+2)$ creation and annihilation operators according to
\begin{equation}
  Q_2^{-1}a_k Q_2 = \sum_{p=k}^{L+1} a_p;\qquad
  Q_2^{-1}a_k^\dag Q_2= a_k^\dagger -a_{k-1}^\dagger (1-\delta_{k,0})
  \label{eq:def_primeda_FP}  
\end{equation}
These relations are analogous to the non-local primed variables we
have introduced for exclusion processes.
%
Moreover, the operator $  {\mathbb H}_3= Q_2^{-1} 
{\mathbb H}_2Q_2$ reads
\begin{eqnarray}
   {\mathbb H}_3  
  &=  
  \sum_{k=0}^{L-1} \big({a^\dag_{k+1}}-{a^\dag_k}\big) 
  \Big\{ 
  W'_{k+1\to k} a_{k+1} - W'_{k\to k+1}a_k
  \Big\}
\end{eqnarray}
This expression corresponds to an \textit{isolated} equilibrium
process of primed rates $W'$, given in~\eref{eq:primed_rates_W}. We
have thus shown that the evolution operator $\mathbb H$ can be mapped
to an equilibrium operator $\mathbb H_3$ acting on a larger space. We
shall now explain how to use this mapping to determine the spectrum of
$\mathbb H$ from that of $\mathbb H_3$.

The operator $ {\mathbb H}_3$ acts on a space of $L+2$ sites ($0\leq
k\leq L+1$) but, as seen from its expression, it only describes
hopping among the first $L+1$ sites ($0\leq k\leq L$), implying that
the last site ($k=L+1$) is completely isolated. Its eigenstates
therefore take the form
\begin{equation}
  |P^{3,\lambda}\rangle = \Big(\bigotimes_{k=0}^{L}|P_k^{3,\lambda}\rangle\Big) \otimes |f^\lambda_{L+1}\rangle
\end{equation}
where $f^\lambda_{L+1}$ is arbitrary. To determine the eigenstates $
|P^\lambda\rangle$ of $\mathbb H$, one maps back the eigenvectors of
$\mathbb H_3$ with $Q_2$ and retains only those which satisfies
\begin{equation}
  \label{eqn:vpcondition}
  Q_2 |P^{3,\lambda}\rangle = |L \rangle \otimes |P^{2,\lambda}\rangle \otimes| R\rangle
\end{equation}
One then has
\begin{equation}
  \label{eqn:construcestate}
  |P^\lambda\rangle = Q_0 Q_1 |P^{2,\lambda}\rangle
\end{equation}

Let us illustrate this procedure to determine the steady-state of
$\mathbb H$. The degenerate ground state of the equilibrium operator
$\mathbb H_3$ takes the form
\begin{equation} 
  | P^{3,0}\rangle
  = \Big( \bigotimes_{k=0}^L | \mathcal{P}_k \rangle\Big )\otimes |f^0_{L+1}\rangle
  \label{eq:def:P1_st}
\end{equation}
where $| \mathcal{P}_k \rangle$ is a Poisson distribution of mean
density $\mu {P_k^\eq}'$, $\mu$ being an arbitrary constant:
\begin{equation}
  |  \mathcal{P}_k \rangle=\sum_{ n_k} e^{-{\mu P_k^\eq}'}  \frac{\big({\mu P_k^\eq}'\big)^{n_k}}{n_k!} |n_k\rangle
\end{equation}
Forcing that $| P^{3,0}\rangle$ satisfies \eref{eqn:vpcondition}
 constrains both $\mu$ and $|f_{L+1}\rangle$:
\begin{equation}
  \mu=\frac{(P_0^\eq)^{-1}-(P_{L+1}^\eq)^{-1}}{\sum_{p=0}^L {P_k^\eq}'};\quad |f_{L+1}\rangle =\sum_{n_{L+1}} e^{{-(P_{L+1}^\eq)^{-1}}}\frac{(P_{L+1}^\eq)^{-n_{L+1}}}{n_{L+1}!} |n_{L+1}\rangle
\end{equation}
One finally checks that applying further $Q_0 Q_1$ to $|
P^{2,0}\rangle$ as in \eref{eqn:construcestate} gives back the steady
state obtained in section \ref{ooop}.

\smallskip

We have thus shown in this section that the evolution operator of
non-interacting particles diffusing on a one-dimensional lattice with
arbitrary rates can be mapped onto an equilibrium operator. The
determination of both steady state and excited states can be mapped to
an equilibrium problem.  The new microscopic feature of the
transformation~\eref{eq:def_primeda_FP} is that the system has been
supplemented with a site at each end, which accounts for the
additional constant of motions of the isolated system ---~the total
mass is conserved.

Last, from the knowledge of the mapping to equilibrium for the
microscopic dynamics, one can extract its counterpart for the
hydrodynamic limit. Indeed we show in
\ref{sec:symmetryFromMicrotoMacro} how similarity transformations for
the evolution operator can be read as canonical changes of variables
in the action. From the expressions of $Q_0$, $Q_1$ and $Q_2$, one
can thus construct a mapping to equilibrium at the level of the action
and then takes its continuum limit, (see section 2.3.5).



\pagebreak

\section{Conclusions}

The macroscopic fluctuation theory, as developed by BDGJL,
 consists of two distinct steps. The first is the recognition that 
the coarse-graining level, as measured by the box size, is  a
parameter playing the role of $\hbar$ in a quantum system:
the dynamics, including large deviations,  becomes in the hydrodynamic limit
a Hamiltonian dynamics.
This step is general, and leaves us with 
a classical field theory to solve.

The second step, that has been followed in a number of systems, is to
calculate explicitly the trajectories  starting from the stationary
situation and ending in an given configuration at long times (i.e.,
an excursion from equilibrium), and hence obtain the large-deviation
function from the action. This can be done trivially in systems
satisfying detailed balance, just by reversing the corresponding
trajectory that relaxes to equilibrium. In systems driven out of
equilibrium, the obvious detailed balance symmetry is broken. For the
driven SSEP and the KMP models, however, a hidden detailed balance symmetry
can be found explicitly, and used to compute the large-deviation
functions.
In this paper we have shown that the reason for this unexpected symmetry,
and hence the solvability, is that these systems can be mapped
back into their equilibrium counterparts.
This realization suggested to look back at the simpler case of independent 
particles diffusing in a potential, with sources at the ends. We have found that
the same strategy can be applied in this elementary case.

The question that arises is how general this mapping is. 
The first problem that comes to mind is how `one-dimensional'  this
mechanism is. One can also ask what happens with more general one-dimensional
cases.  A way to start investigating the first question is to consider the
simple case of non-interacting particles, but in higher dimensions.
On the other hand,  more general one-dimensional models might perhaps admit 
 a mapping back to equilibrium if one is prepared
to pay the price of dealing with spatially non-local interactions.

\vspace{.5cm}

{\bf Acknowledgments}

We wish to thank H. Fogedby and H. Spohn for useful discussions. JT
acknowledges funding from EPSRC grants EP/030173 and GR/T11753.

\pagebreak

\appendix

\section{Coherent state path-integral}

\label{app:action_func}

The functional expression for spin operators presents a few mathematical 
subtleties~\cite{Garg2000}.  Recently 
Solari~\cite{Solari1987}, Kochetov~\cite{Kochetov1995}, Vieira and
Sacramento~\cite{Vieira1995} derived an expression for the
action and  the associated time boundary conditions.  We follow here
the clear presentation of Stone, Park and Garg~\cite{Garg2000}.

Usually, one seeks to calculate the  `propagator' between two normalized
coherent states~\cite{Kochetov1995}
\begin{equation}
 \langle {\bol z}^f| \rme^{-T \hat H}| {\bol z}^i\rangle
\end{equation}
As  usual in the  construction of a path integral, the time interval
$[0,T]$ is divided into $N$ segments and one inserts at each time
interval $L$ representation of the identity~\eref{eq:represid}, one
for each site of the lattice. Letting $N$ go to infinity, one ends up
with the functional representation of the propagator
\begin{eqnarray}
  \label{eq:propzizf}
 \langle {\bol z}^f| \rme^{-T \hat H}| {\bol z}^i\rangle=
  \int\D\bar {\bol z}\D {\bol z} \exp[-\tilde S]\\
  \tilde S = - j \sum_k
  \log \frac{ (1+\bar z_k^f z_k(T))(1+\bar z_k(0) z_k^i) }{(1+\bar
  z_k^f z_k^f)(1+\bar z_k^i z_k^i)} + 2j \int_0^T \rmd t\: \left[
  \frac 12\sum_k \frac{\bar z_k \dot z_k - \dot {\bar z}_k z_k}{1+\bar
  z_k z_k} - \HH(\bol{ \bar z},\bol{ z})\right]\label{eqn:zpropaaction}
\end{eqnarray}
Note that the temporal boundary term differs slightly from those
in~\cite{Kochetov1995,Garg2000} as the representation we used of the
SU(2) group is not the usual unitary one (cfr
\ref{app:UnitorNonUnit}). The role of the Hamiltonian is played by the
quantity
\begin{equation}
  \label{eq:defWscalar}
  \HH(\bol{ \bar z},\bol{ z}) = -\frac 1{2j} \langle \bol{ z}|\hat H |
  \bol{ z}\rangle
\end{equation}
which is computed using~\eref{eq:classicS} or similar expressions for
higher powers of the spin operators (see~\cite{Garg2000}).
 $\bar z_k(t)$ and $z_k(t)$ are two independent complex
fields~\cite{Solari1987}. In the construction of the path integral,
one notes that $z_k(0)=z_k^i$ and $\bar z_k(T) = \bar
z_k^f$~\cite{foot1}, while $z_k(T)$ and $\bar z_k(0)$ are
unconstrained~\cite{Solari1987, Kochetov1995}.

Here, we rather wish to compute the physical propagator between two
states $\langle \bol{ n^f}|$ and $|\bol{ n^i} \rangle$, corresponding
to fixed initial and final number of particles in each site. The
propagator represents the probability $P(\bol{ n^f},T;\bol{ n^i},0)$
of observing the system in state $(n_1^f,\dots,n_L^f)$ at time $T$,
starting from $(n_1^i,\dots,n_L^i)$ at time $0$. Using $2L$
representations of identity~\eref{eq:represid}, we write
\begin{eqnarray}
  \label{eqn:physpropa}
  P(\bol{ n^f},T;\bol{ n^i},0)
  &=\langle \bol{ n}^f| \rme^{-T \hat H}| \bol{ n}^i\rangle 
  =\int\prod_k \rmd\mu(z_k^f)\rmd\mu(z_k^i)\:
  \langle \bol{ n}^f|\bol{ z}^f\rangle 
  \langle \bol{ z}^f| \rme^{-T \hat H}| \bol{ z}^i\rangle
  \langle \bol{ z}^i|\bol{ n}^i\rangle 
\end{eqnarray}
Keeping in mind that we are aiming to describe a hydrodynamic limit,
we introduce the discrete densities 
\begin{equation}
  \rho_k=n_k/2j
\end{equation}
One sees by comparing equations \eref{eq:propzizf} and
\eref{eqn:physpropa} that  the physical propagator is obtained by
subtracting $\log \langle \bol{ z}^i|\bol{ n}^i\rangle + \log \langle \bol{
n}^f|\bol{ z}^f\rangle$ to the action \eref{eqn:zpropaaction}, and by
integrating over $z^i,\bar z^i, z^f,\bar z^f$. As we aim to describe
the large $j,L$ limit, we can use Stirling's formula to obtain the
asymptotics of $\smallmatrix{ 2j\cr n_k^f}\!\!$, to get:
\begin{eqnarray}
  \label{eq:propninfapp}
  P(\bol{ \rho}^f,T;\bol{ \rho}^i,0)= \int\prod_k
   \rmd\mu(z_k^f)\rmd\mu(z_k^i)\int\D\bol{ \bar z}\D \bol{ z}
   \exp[-S]\\ S = - j\sum_k\log [(1+\bar
   z_k^f z_k(T)) (1+\bar z_k(0) z_k^i)] + 2j\int_0^T \rmd t\: \left[
   \frac 12\sum_k \frac{\bar z_k \dot z_k - \dot {\bar z}_k
   z_k}{1+\bar z_k z_k} - \HH(\bol{\bar z},{\bol z})\right]\nonumber
   \\ +2j\sum_k \label{eq:actioninf}\left\{ -\rho_k^i \log \bar z_k^i -
   \rho_k^f \log z_k^f + \rho_k^f \log \rho_k^f + (1-\rho_k^f)\log
   (1-\rho_k^f) +  \log[(1+\bar z_k^f
   z_k^f)(1+\bar z_k^i z_k^i)]\right\}\nonumber
\end{eqnarray}
The complex fields $z_k^i,\bar z_k^i, z_k^f$ and $\bar z_k^f$ have now to be
integrated upon, which can be done by saddle point. First
differentiating the action~\eref{eq:propninfapp} with respect to $\bar
z_k^i$, $z_k^f$ yields
\begin{eqnarray}
\frac{1}{2j} \frac{\partial S}{\partial \bar z_k^i}=
  \frac{z_k^i}{1+\bar z_k^iz_k^i} - \frac{\rho_k^i}{\bar z_k^i} \qquad
  \frac{1}{2j} \frac{\partial S}{\partial z_k^f}= \frac{\bar
  z_k^f}{1+\bar z_k^fz_k^f} - \frac{\rho_k^f}{ z_k^f}
\end{eqnarray}
We thus obtain the initial and final conditions
\begin{eqnarray} \label{eq:timeboundaryzzapp}
  \frac{\bar z_k^iz_k^i}{1+\bar z_k^iz_k^i} = \rho_k^i  \qquad
  \frac{\bar z_k^fz_k^f}{1+\bar z_k^fz_k^f} = \rho_k^f
\end{eqnarray}
Extremalizing with respect to $z_k^i$ and $\bar z_k^f$ has to be done
carefully as the time integral in the action gives a non-zero
contribution (See for instance equation 3.9 of~\cite{Garg2000}). Such
a computation leads to
\begin{eqnarray}
\frac{1}{2j} \frac{\partial S}{\partial z_k^i}= \frac{\bar
  z_k^i}{1+\bar z_k^iz_k^i} - \frac{\bar z_k(0)}{1+\bar z_k(0)z_k^i}
\qquad \frac{1}{2j} \frac{\partial S}{\partial \bar z_k^f}= \frac{
  z_k^f}{1+\bar z_k^fz_k^f} - \frac{z_k(T)}{1+ z_k(T)\bar z_k^f}
\end{eqnarray}
which fixes
\begin{equation}
  \label{eq:timeboundaryzzapp2}
  \bar z_k(0) = \bar z_k^i \qquad z_k(T) = z_k^f
\end{equation}
Putting everything together, the action reads
\begin{eqnarray}
  S[\bol{ \bar z},{\bol z};\bol{ n_f},\bol{ n_i}] &= 2j \sum_k
  \left[\frac{z_k \bar z_k}{1+z_k \bar z_k}\log\bar z_k - \log(1+z_k
    \bar z_k)\right]_i^f +2 j \int \rmd t \Big[\sum_k \frac{\bar z_k \dot
      z_k}{1+\bar z_k z_k} -\HH(\bar z,z)\Big]\nonumber
  \end{eqnarray}
\begin{equation}
\HH_B(\bar z,z)=- \frac 12 \sum_{k=1}^{L-1} \frac{ z_k z_{k+1} \, (\bar z_{k+1}-\bar
    z_k)^2 }{(1+\bar z_k z_k)(1+\bar z_{k+1} z_{k+1})}  + \left(\frac { z_{k+1}} {1+z_{k+1} \bar z_{k+1}}-\frac
  { z_k} {1+z_k \bar z_k}\right)({\bar z_{k+1}-\bar z_k})
\label{eqn:hamiltonianzapp}
\end{equation}

In the context of quantum mechanics, the coherent state
(stereographic) parametrization in terms of $z_k(t)$ and $\bar z_k(t)$
is usually transformed into spherical polar coordinates through
$z_k=e^{-i\phi_k} \cot \frac {\theta_k} 2$, $\bar z_k=e^{i\phi_k} \cot
\frac {\theta_k} 2$. In our context however, it will prove more
convenient to introduce a new parametrization \cite{foot2}
\begin{equation}
  z_k = \frac{\rho_k}{1-\rho_k}\,e^{-\hat \rho_k} \:,\qquad
  \bar z_k = e^{\hat \rho_k}
\end{equation}
Using $\frac{\bar z_kz_k}{1+\bar z_kz_k}=\rho_k$ we see
from~(\ref{eq:timeboundaryzzapp},\ref{eq:timeboundaryzzapp2}) that the temporal
boundary conditions on the fields can be written as
\begin{equation}
  {\bol\rho}(0)=\bol{ \rho^i} \:, \qquad {\bol \rho}(T)=\bol{ \rho^f}
  \:, \qquad \textnormal{with} \quad \bol{ \hat\rho}(0),\: \bol{
    \hat\rho}(T)\ \textnormal{unconstrained}
\end{equation}
This highlights the correspondence between the field ${\bol \rho}$ and
the actual density of the system. The Hamiltonian
\eref{eqn:hamiltonianzapp} then reads
\begin{eqnarray}
  \label{eqn:microham}
  \HH=\HH_B+\HH_0+\HH_L,\\
  \HH_B=\frac 1 2\sum_{k=1}^{L-1}\left\{ 
  \,(1-\rho_k) \rho_{k+1} \left[\rme^{\hat \rho_k - \hat
  \rho_{k+1}}-1\right]+  \,\rho_k
  (1-\rho_{k+1})\left[\rme^{\hat \rho_{k+1}-\hat
  \rho_k}-1\right]\right\}\\ \HH_0=\alpha(1-\rho_1)(e^{\hat
  \rho_1}-1)+\gamma \rho_1(\rme^{-\hat\rho_1}-1)\\
  \HH_L=\delta
  (1-\rho_L)(\rme^{\hat\rho_L}-1)+\beta \rho_L
  (\rme^{-\hat\rho_L}-1)
\end{eqnarray}
where $\HH_B$ describes the interaction in the bulk whereas $\HH_0$
and $\HH_L$ result from the coupling to the reservoirs. Using
\begin{equation}
  \frac 12
  \frac{\bar z_k\dot z_k -\dot{\bar z}_kz_k}{1+\bar z_kz_k}= \dot\rho_k
  \hat\rho_k-\partial_t\big[\hat\rho_k\rho_k+\frac
  12\log(1-\rho_k)\big]
\end{equation}
we check that all boundary terms cancel so that the classical action
reduces to
\begin{equation}
  \label{eq:appaction1site}
  S[\hat\rho,\rho;\rho_i,\rho_f]= 2j \int_0^T \rmd t \left[ \sum_k
    \hat\rho_k \dot\rho_k - \HH(\hat \rho,\rho)\right]
\end{equation}

\section{Hydrodynamic limit}
\label{app:hydrolim}
\subsection{Continuous limit of the action}

Let us start from the microscopic Hamiltonian \eref{eqn:microham} for
a system of size $L$. In the large $j$ limit, each site contains a
large number of particles, so that the density field tends to
self-average. It is thus natural to assume the gradients to be small.
Expanding the Hamiltonian up to second order in $(\rho_{k+1}-\rho_k),\
(\hat\rho_{k+1},\hat \rho_k)$, one gets
\begin{equation}
  \HH_B=\sum_{k=1}^{L-1}\left[ \frac 1 2 \, \rho_k (1-\rho_k)
  (\hat\rho_{k+1}-\hat\rho_k)^2 - \frac 1
  2(\hat\rho_{k+1}-\hat\rho_k)(\rho_{k+1}-\rho_k)\right]
\end{equation}
One can then introduce a rescaled space variable
\begin{equation}
  x_k=\frac{k}{2j}
\end{equation}
which goes from $1/2j$ to $L/2j$ and becomes continuous in the large j
limit. Assuming a diffusive time scale and explicitly rescaling the
gradients
\begin{equation}
  \rho_{k+1}-\rho_k\to \frac 1 {2j} \grad \rho,\quad
  \hat\rho_{k+1}-\hat\rho_k\to \frac 1 {2j} \grad \hat\rho,\quad \frac
  1 {2j} \sum_{k=1}^L \to \int_0^{L/2j} \rmd x,\quad \rmd t \to (2j)^2
  \rmd t,\quad
\end{equation}
transforms the action into
\begin{eqnarray}
  S[\rho(x),\hat\rho(x)]=(2j)^2 \int \rmd t \left\{\int_0^{L/2j}\!\!\!\! \rmd
  x \left[ \hat\rho \dot \rho - \frac 1 2 \sigma (\grad \hat \rho)^2 +
  \frac 1 2 \grad \rho \grad \hat \rho \right]-2j\,
  \HH_0[\rho,\hat\rho]-2j\, \HH_1[\rho,\hat\rho] \right\}\nonumber\\
  \HH_0[ \bol \rho , \bol{ \hat \rho}]= \alpha\,[1-\rho(0)](e^{\hat
  \rho(0)}-1)+ \gamma \,\rho(0)(\rme^{-\hat\rho(0)}-1)\\ \HH_1[ \bol
  \rho , \bol{ \hat \rho}]=\nonumber \delta
  \,[1-\rho(1)](\rme^{\hat\rho(1)}-1)+ \beta \,\rho(1)
  (\rme^{-\hat\rho(1)}-1)
\end{eqnarray}
It is however more convenient to have a continuous variable $x$ going
from 0 to 1, so that the extensivity of the action is explicit. This
can be achieved by a further scaling
\begin{equation}
  x \to \frac L{2j}\, x,\quad t \to \left(\frac L{2j}\right)^2\,t
\end{equation}
which maps the action into
\begin{eqnarray}
  S[\rho(x),\hat\rho(x)]=2j\,L \!\!\!\int\!\!\! \rmd t \left\{\int_0^1 \rmd x
  \left[ \hat\rho \dot \rho - \frac 1 2 \sigma (\grad \hat \rho)^2 +
  \frac 1 2 \grad \rho \grad \hat \rho \right]-L\,
  \HH_0[\rho,\hat\rho]-L\, \HH_1[\rho,\hat\rho] \right\}    \label{eq:continuumaction}
\end{eqnarray}
Some comments are in order. First, the bulk integral is proportional
to $2j L$, which is consistent with the scaling of a large deviation
function. Then, the boundary terms scale as $2j L^2$ and deserve
some further analysis.

\subsection{Spatial boundary conditions}
\label{app:BC}

Let us first note that if the boundary rates $\alpha, \beta, \gamma,
\delta$ are of order $1/L$, then the contributions of $\HH_0$ and
$\HH_L$ to the action are of the same order as that of the bulk term
$\HH_B$. This means that trajectories with fluctuations of order 1 at
the boundaries give rise to non-vanishing contribution at the level of
large deviation, i.e. their action scales as $2j L$. In particular,
such rates would {\em not} lead to the spatial boundary conditions
\eref{eqn:BC1}.

In the usual case where the rates are of order 1, we shall show below
that trajectories with fluctuations at the boundaries are forbidden at
the level of large deviations. This accounts for the claim
in~\cite{Bertini2001,Bertini2005} that trajectories which do not
satisfy strictly the boundary condition correspond to infinite values
of the large deviation function.

Let us analyze in details the dynamics at the boundaries. In the large
$j$ limit, the probability is dominated by trajectories that
extremalize the action (see section \ref{sec:LDLNL}). Using
expression \eref{eqn:microham} for the microscopic Hamiltonian, the
classical equations read:

\begin{eqnarray}
\label{eqn:motionsbound}
  \dot \rho_1=\frac{\partial (\HH_1+\HH_B)}{\partial
    \hat\rho_1}=\frac{1-\rho_1}2\rho_2 e^{\hat\rho_1-\hat\rho_2}-\frac
    {\rho_1}2(1-\rho_2)e^{\hat\rho_2-\hat\rho_1}+\alpha(1-\rho_1)
    e^{\hat\rho_1}-\gamma \rho_1 e^{-\hat\rho_1}\\ 
\dot {\hat \rho}_1=-\frac{\partial (\HH_1+\HH_B)}{\partial
    \rho_1}=\frac {\rho_2 }2[e^{\hat\rho_1-\hat\rho_2}-1]-
    \frac{1-\rho_2}2[e^{\hat\rho_2-\hat\rho_1}-1]+\alpha
	 [\e^{\hat\rho_1}-1]-\gamma [e^{-\hat\rho_1}-1]\nonumber
\end{eqnarray}
In the large $2j L$ limit, one checks that $\partial_{\rho_1} \HH_B$
and $\partial_{\hat \rho_1}\HH_B$ are of order $1/L$. The r.h.s. of
equations \eref{eqn:motionsbound} are thus dominated by the boundary
terms so that, at first order, they read
\begin{eqnarray}
  \label{eqn:motionsboundapp}
  \dot \rho_1=\alpha(1-\rho_1) e^{\hat\rho_1}-\gamma \rho_1
  e^{-\hat\rho_1}\\ 
  \dot {\hat \rho}_1=\alpha
  [\e^{\hat\rho_1}-1]-\gamma [e^{-\hat\rho_1}-1]\nonumber
\end{eqnarray}
For sake of clarity, we now drop the index `$1$'. Equations
\eref{eqn:motionsboundapp} can be solved and give
\begin{eqnarray}
  \rho(t)=\left[\rho_0 (1-\rho_i)(1-e^{-\Gamma
      t})+\rho_ie^{-\hat\rho_i}(\rho_0+(1-\rho_0)e^{-\Gamma
      t})\right]\times\nonumber\\
  \label{eqn:solrho1rhohat1}
  \qquad\qquad\left[(1-\rho_0)(1-e^{\Gamma t})+e^{\hat\rho_i}(\rho_0+(1-\rho_0)e^{\Gamma t})\right]\\
  e^{\hat\rho(t)}=\frac{\rho_0 e^{\hat \rho_i}+1-\rho_0+(1-\rho_0)e^{\Gamma t}(e^{\hat\rho_i}-1)}{\rho_0 e^{\hat \rho_i}+1-\rho_0-\rho_0 e^{\Gamma t}(e^{\hat\rho_i}-1)}
\end{eqnarray}
where $\Gamma=\alpha+\gamma$, $\rho_0=\alpha/\Gamma$ and $\rho_i,\hat
\rho_i$ are the initial conditions of the fields. For such a trajectory,
the contribution $S_L$ of the left boundary to the action is
\begin{equation}
  S_L=-2 j t \Gamma (1-e^{-\hat \rho_i})\left[\rho_0 (1-\rho_i)e^{\hat \rho_i}-\rho_i (1-\rho_0)\right]
\end{equation}
In the hydrodynamic limit, the time is rescaled by $L^2$ so that $S_L
\sim 2 j L^2$. We know from section \ref{sec:LDFfromACtion} that the
probability to observe a given profile is obtained from the
exponential of the action. For non-zero $S_L$, it is thus of order
$\exp(-2j L^2)$ and such profile is even more rare than large
deviations, whose probability scale as $\exp(-2j L)$. Quantitatively,
such profile has an infinite large deviation function
\begin{equation}
  {\cal F}[\rho]= -\lim_{2j L \to \infty}\frac 1 {2j L} \log P[\rho]=\lim_{2j L \to \infty}\frac 1 {2j L} S_{\text{traj}}=\infty
\end{equation}

In addition to an infinite large deviation function, the trajectories
\eref{eqn:solrho1rhohat1} can also present a diverging density. Indeed,
for large times, one gets
\begin{equation}
  \rho(t) \sim e^{\Gamma t}\rho_0 (1-\rho_0)(1-e^{-\hat\rho_i}) [\rho_i+(1-\rho_i)e^{\hat\rho_i}]
\end{equation}
To keep the density finite, one thus needs
\begin{equation}
  \hat \rho_i=0\qquad\text{or}\qquad \rho_i=\frac{1}{1-e^{-\hat\rho_i}}
\end{equation}
The second solution corresponds to an action $S_L=2jt \Gamma$ which
is once again of order $2j L^2$ in the hydrodynamic limit. Such a
trajectory is thus forbidden at the level of large deviations. We are
left with $\hat \rho_i=0$, which corresponds to $\rho(t)=
\rho_0+e^{-\Gamma t} (\rho_i-\rho_0)$. The corresponding contribution
$S_L$ vanishes exactly and in a macroscopic time of order $L^{-2}$ the
density $\rho(t)$ of the first site gets equal to $\rho_0=\frac
\alpha{\alpha+\gamma}$.

The same analysis holds for the right boundary and we have thus shown
that for boundary rates of order 1, the fields $\rho, \hat \rho$ have
to satisfy the following boundary conditions at the level of large
deviations:
\begin{equation}
  \hat \rho(0,t)=\hat\rho(1,t)=0;\qquad \rho(0,t)=\rho_0=\frac\alpha{\alpha+\gamma};\qquad\rho(1,t)=\rho_1=\frac\delta{\delta+\beta}
\end{equation}
whereas for rates of order $1/L$, fluctuations are allowed at the
level of large deviations.

\section{Unitary and non-unitary representations}
\label{app:UnitorNonUnit}
\subsection{Change of basis}
We wish to make the link between the representation of $SU(2)$
matrices we have used \eref{eq:spin_matrx} and the usual ``quantum''
representation.  In terms of the action of these operators, we
 start from the action of the matrices~\eref{eq:spin_matrx} on
the occupation kets~$|n\rangle$
\begin{eqnarray}
  S^+|n\rangle &= (2j-n)|n+1\rangle \\
  S^-|n\rangle &= n|n-1\rangle\\
  S^z|n\rangle &= (n-j)|n\rangle
\end{eqnarray}
and perform a similarity transformation to obtain the canonical
unitary representation of ``quantum'' operators, that we denote
$S_q^\pm$, $S_q^z$:
\begin{eqnarray}
   Q^{-1}S^+Q|n\rangle= S_q^+|n\rangle &=& \sqrt{(n+1)(2j-n)}
   |n+1\rangle \label{eq:eq_q_p}\\ Q^{-1}S^-Q|n\rangle= S_q^-|n\rangle
   &=& \sqrt{n(2j-n+1)}
   |n-1\rangle \label{eq:eq_q_m}\\ Q^{-1}S^zQ|n\rangle= S_q^z|n\rangle
   &=& (n-j) |n\rangle \label{eq:eq_q_z}
\end{eqnarray}
The usual magnetic number $m$ is related to the occupation number $n$
through $m=n-j$.  We first remark that only $S^+$ and $S^-$ have to
be changed. This suggests we find $Q$ as a function of the number
operator $\hat n=j+S^z$:
\begin{equation}
  Q=q(\hat n)
\end{equation}
From~(\ref{eq:eq_q_p}-\ref{eq:eq_q_m}) we have to solve
\begin{eqnarray}
  n \frac{q(n)}{q(n-1)} &=&  \sqrt{n(2j-n+1)} \\
  (2j-n) \frac{q(n)}{q(n+1)} &=&\sqrt{(n+1)(2j-n)}
\end{eqnarray}
which is done by choosing
\begin{equation}
  Q^{-1} = \sqrt{\smallmatrix{2j\cr \hat n}}
\end{equation}

\medskip
\subsection{Terms arising from the similarity transformation}

\label{app:newbasis}

\medskip  We now denote the  basis $|n\rangle$ of
kets for occupation numbers as
$|\theta\rangle$, with $j\cos\theta_k=2n_k-1$ on each site.  We are
interested in the propagator
\begin{equation}\label{eq:propagat}
  P(n',t|n,0)=\langle n'|e^{-t\hat H}|n\rangle
\end{equation}
where $\hat H$ is expressed in terms of the non-unitary
representation~\eref{eq:spin_matrx}.  However, the field-theoretic
construction of the action corresponding to~\eref{eq:propagat} is well
suited only for `quantum' spin operators acting on kets as
in~(\ref{eq:eq_q_p}-\ref{eq:eq_q_z}).  To make the bridge between
these representations, let us write
\begin{equation}
  P(n',t|n,0)=\langle n'|Q e^{-t\hat H_q} Q^{-1}|n\rangle
\end{equation}
where $\hat H_q$ is now expressed in terms of `quantum' spin matrices,
that is to say, in terms of the matrices $S_q^{+,-,z}$, 
which act on $|n\rangle$
according to~(\ref{eq:eq_q_p}-\ref{eq:eq_q_z}).  Inserting two
resolution of the identity yields
\begin{equation}
   P(n',t|n,0)=\int d\xi_fd\bar\xi_f  d\xi_id\bar\xi_i 
   e^{-S(\theta',\theta)}
\end{equation}
with
\begin{equation}
  S(\theta',\theta) =
  -\ln \frac{\langle n'|Q|\xi_f\rangle}{\langle \xi_f|\xi_f\rangle}
  -\ln \frac{\langle \xi_i|Q^{-1}|n\rangle}{\langle \xi_i|\xi_i\rangle}+
  S(\xi_i,\bar\xi_f)
\end{equation}
But $Q$ is diagonal and real-valued on the basis $|n\rangle$ so that
\begin{equation}
  S(\theta',\theta) =
  -\ln \frac{\langle \theta'|\xi_f\rangle}{\langle \xi_f|\xi_f\rangle}
  -\ln \frac{\langle \xi_i|\theta\rangle}{\langle \xi_i|\xi_i\rangle}+
  \frac 1 2 \ln \frac{\smallmatrix{2j\cr n}}{\smallmatrix{2j\cr n'}}+
  S(\xi_i,\bar\xi_f)
  \label{eq:SunitSnonunit}
\end{equation}
We thus conclude that using the non-unitary representation of $SU(2)$
yields a new term in the action (the ratio of binomials
in~\eref{eq:SunitSnonunit}), that we took into account to
obtain~\eref{eqn:zpropaaction}.

\section{From microscopic detailed-balance 
relation to the symmetry of the action in the path integral}
\label{sec:symmetryFromMicrotoMacro}

Let us first discuss in 
detail the relation between similarity transformations of
the evolution operator and canonical changes of variable in the
action, and then show the consequences of the existence of symmetries
for the system. We then analyze the case of the detailed balance
relation. To remain in the context of this article, we present
below the case of the SSEP in the hydrodynamic limit although the results
only rely on the existence of the path-integral representation and
are thus much more general.

We consider the propagator between two physical states
\begin{equation}
  \label{eqn:propag}
  G(\bol \rho_f,\bol \rho_i;T)=\langle \bol \rho_f | e^{-T H} | \bol \rho_i \rangle
\end{equation}
where $H$ is the evolution operator. We know from section
\ref{sec:spin_coh_st} that the path integral representation of this
propagator is given by
\begin{equation}
  \label{eqn:actionS}
  G(\bol \rho_f,\bol \rho_i;T)=\int {\cal D}[\bol {\hat \rho},
  \bol\rho] e^{-2j L S_H};\qquad S_H=\int dt dx \{\hat \rho\dot \rho -
  \HH[\rho,\hat \rho]\}
\end{equation}
where $\HH[\rho,\hat \rho]=-\frac 1 {2j}\langle z| H | z\rangle$ and
we used the relations \eref{eqn:transfo} to introduce densities. Let
us introduce the operator obtained after a similarity transformation
\begin{equation}
  \tilde H = Q^{-1} H Q
\end{equation}
\eref{eqn:propag} can then  be written
\begin{equation}
  \label{eqn:propag2}
  G(\rho_f,\rho_i;T)=\langle \rho_f | Q Q^{-1} e^{-T H} Q Q^{-1} | \rho_i \rangle=\langle \rho_f | Q  e^{-T \tilde H} Q  | \rho_i \rangle
\end{equation}
The path-integral representation of \eref{eqn:propag2} leads to 
\begin{equation}
  G(\rho_f,\rho_i;T)=\int {\cal D}[\hat \rho', \rho'] e^{-2jL S_{\tilde H}};\qquad
  S_{\tilde H}=-\frac 1 {2jL} [\log {\cal Q}]^f_i+\int dt dx \{\hat \rho'\dot \rho' - \tilde \HH[\rho',\hat \rho']\}
  \label{eqn:actionShat}
\end{equation}
where $\tilde \HH[\rho',\hat \rho']=-\frac{1}{2j}\langle z| \tilde H | z\rangle$ and ${\cal
Q}=\langle z| Q | z\rangle$. It is now simple to check that the
actions \eref{eqn:actionS} and \eref{eqn:actionShat} are related via
the canonical changes of variables induced by $\log {\cal Q}$~\cite{Goldstein1980}:
\begin{equation}
  \label{eqn:cantrans}
  \int dx \dot \rho' \hat \rho'= \frac 1 {2jL}\frac{d }{dt}\log{\cal Q} + \int dx
  \dot \rho \hat \rho
\end{equation}

If $Q$ is a symmetry of the evolution operator, we see that the
commutation relation $[H,Q]=0$ can also be written $H= Q^{-1} H Q$.
At the level of the action, it thus means that $\log {\cal Q}$ induces
a canonical transformation that leaves the action invariant up to
boundary terms.

The case of detailed balance relation is slightly different as it is
not a symmetry of the evolution operator but rather a connection
between $H$ and its adjoint $H^\dagger$, as we recall. Indeed,
for any configuration $\bol \rho_1$, $\bol \rho_2$ it reads
\begin{equation}
  \langle \bol \rho_1| e^{-tH} | \bol \rho_2 \rangle P_\eq(\bol \rho_2) =   \langle \bol \rho_2| e^{-tH} | \bol \rho_1 \rangle P_\eq(\bol \rho_1) 
\end{equation}
Taking the adjoint of the r.h.s, one gets
\begin{equation}
  \langle \bol \rho_1 |e^{-tH} P_{\eq}-P_{\eq} e^{-tH^\dagger} |\bol \rho_2  \rangle
\end{equation}
As this holds for all $\bol \rho_1, \bol \rho_2,t$ we see by deriving with
respect to t and putting $t=0$ that 
\begin{equation}
  H^\dagger= P_{\eq}^{-1} H P_{\eq}
\end{equation}
At the level of the actions, it then reads
\begin{equation}
  \int \hat\rho \dot \rho - \HH[\rho,\hat \rho]=-\frac{1}{2j L} [\log
  {\cal P}_\eq]_i^f + \int \hat\rho \dot \rho - \HH^\dagger[\rho,\hat
  \rho]
\end{equation}
In the case of the SSEP with periodic boundary conditions, one further knows 
that
\begin{equation}
  P_\eq[\rho]\propto e^{-2j L \int dx \rho \log \rho+(1-\rho)\log(1-\rho)}
\end{equation}
and
\begin{equation}
  \HH[\rho,\hat \rho]=\frac 1 2 \sigma \grad \hat \rho^2 - \frac 1 2 \grad \rho \grad \hat \rho;\qquad  \HH^\dagger[\rho',\hat\rho']=\frac 1 2 \sigma \grad \hat {\rho'}^2 + \frac 1 2 \grad \rho' \grad \hat \rho';
\end{equation}
We thus see that the action of $H$ and $H^\dagger$ are related
through the canonical changes of variable induced by
\eref{eqn:cantrans}:
\begin{equation}
  \rho'=\rho;\qquad \hat \rho'=\hat\rho-\frac{\rho}{1-\rho}
\end{equation}
By further taking a time reversal transformation
\begin{equation}
  t\to T-t;\qquad \hat \rho' \to -\hat \rho'
\end{equation}
one maps back $\HH^\dagger$ to $\HH$ and thus obtain a symmetry of the
action, which finally reads
\begin{equation}
  \rho_{\mbox{\tiny TR}}(t)=\rho(T-t);\qquad \hat \rho_{\mbox{\tiny TR}}(t)= -\hat \rho(T-t) +\frac
       {\rho(T-t)}{1-\rho(T-t)}
\end{equation}
We get back the transformation \eref{eqn:transfoDBrhos}, as expected.

\pagebreak


\begin{thebibliography}{99}
\bibitem{foot1} These fields coincide exactly, not as the result of a
  saddle point evaluation.
\bibitem{foot2} In the changes of variable $z,\bar z\to \rho,\hat
  \rho$, the Jacobian $(1-\rho)^2$ cancel exactly the prefactor of
  $\rmd \mu(z)$, so that $\int \rmd \mu(z)$ is replaced by $\int
  \frac{2j+1} \pi \rmd \rho \rmd \hat \rho$.
\bibitem{foot4a} Note that with this convention the spin normalization
  is $ \bol S \cdot \bol S=\frac 14$ instead of the usual $ \bol S
  \cdot \bol S=1$.
\bibitem{foot4} This can be checked using the expressions
  \eref{eq:classicSrho} in the continuum space limit and the Poisson
  bracket $\{A(x),B(y)\}= \int \text{d}z \big\{ \frac{\delta
    A(x)}{\delta \rho(z)} \frac{\delta B(y)}{\delta \hat\rho(z)} -
  \frac{\delta B(y)}{\delta \rho(z)} \frac{\delta A(x)}{\delta
    \hat\rho(z)} \big\}$.
\end{thebibliography}
\end{document}